\documentclass[12pt]{article}

\usepackage[parfill]{parskip}
\usepackage[utf8]{inputenc}
\usepackage{natbib}
\usepackage{amsmath,amssymb,amsfonts,amsthm}
\usepackage{subfiles}
\usepackage{algorithm,algorithmic}

\newcommand{\No}{ \text{No} }

\newcommand{\tr}{ \text{tr} }
\newcommand{\etr}{ \text{etr} }
\usepackage{bbold}

\newcommand{\diag}[1]{ \text{diag}\{#1\}}

\newcommand{\bel}[1]{\begin{equation}\label{#1}\begin{aligned}}

\newcommand{\eel}{\end{aligned}\end{equation}}

\newcommand{\sigmae}{\sigma^2_e}

\newcommand{\be}{\begin{equation*}\begin{aligned}}

\newcommand{\ee}{\end{aligned}\end{equation*}}
\newtheorem{theorem}{Theorem}
\newtheorem{remark}{Remark}
\newtheorem{lemma}{Lemma}

\usepackage{graphicx}

\usepackage{subcaption}
\usepackage{float}

\newcommand{\blind}{1}

\begin{document}

\def\spacingset#1{\renewcommand{\baselinestretch}%
{#1}\small\normalsize} \spacingset{1}


\if1\blind
{
  \title{\bf  Spiked Laplacian Graphs:\\
 Bayesian Community Detection in Heterogeneous Networks}
  \author{Leo L. Duan \thanks{Department of Statistics, University
of Florida, Gainesville, FL, email: li.duan@ufl.edu},
George Michailidis \thanks{Department of Statistics \& UF Informatics Institute, University
of Florida, Gainesville, FL, email: gmichail@ufl.edu},
Mingzhou Ding \thanks{J. Crayton Pruitt Family  Department of Biomedical Engineering,
University
of Florida, Gainesville, FL, email: mding@bme.ufl.edu}
}
   \date{}
  \maketitle
} \fi

\if0\blind
{
  \bigskip
  \bigskip
  \bigskip
  \begin{center}
    {\bf  Spiked Laplacian Graphs:\\
 Bayesian Community Detection in Heterogeneous Networks}
\end{center}
} \fi

\bigskip
\begin{abstract}
In network data analysis, it is becoming common to work with a collection of graphs that exhibit \emph{heterogeneity}. For example, neuroimaging data from patient cohorts are increasingly available. A critical analytical task is to identify communities, and graph Laplacian-based methods are routinely used. However, these methods are currently limited to a single network and do not provide measures of uncertainty on the community assignment. In this work, we propose a probabilistic network model called the ``Spiked Laplacian Graph'' that considers each network as an invertible transform of the Laplacian, with its eigenvalues modeled by a modified spiked structure. This effectively reduces the number of parameters in the eigenvectors, and their sign patterns allow efficient estimation of the community structure. Further, the posterior distribution of the eigenvectors provides uncertainty quantification for the community estimates. Subsequently, we introduce a Bayesian non-parametric approach to address the issue of heterogeneity in a collection of graphs. Theoretical results are established on the posterior consistency of the procedure and provide insights on the trade-off between model resolution and accuracy. We illustrate the performance of the methodology on synthetic data sets, as well as a neuroscience study related to brain activity in working memory.
\end{abstract}
\noindent%
\vfill

\spacingset{1.5} 

\section{Introduction}

In recent years, there has been a strong interest in modeling network data due to their increased availability in social sciences \citep{aggarwal2011introduction}, biology \citep{minch2015dna} and engineering \citep{zhang2008global}. A popular generative model, suitable for social network analysis has been the stochastic block model \citep{nowicki2001estimation,karrer2011stochastic} and its variant, the mixed membership stochastic block model \citep{airoldi2008mixed}. Their applicability stems from the fact that these models tend to produce networks organized in communities; subsets of vertices connected with one another with particular edge densities. For example, edge density may be higher within communities than between communities. A key analytical task is that of community detection and a plethora of
algorithms have been proposed in the literature [\cite{leighton1999multicommodity,khandekar2009graph,arora2009expander,mucha2010community,fortunato2010community, papadopoulos2012community};
for recent reviews, see \cite{abbe2017community,javed2018community}].
Further, consistency results when the number of vertices grows to infinity have also been provided for certain community detection algorithms, with spectral clustering being the most prominent among them \citep{rohe2011spectral,amini2013pseudo}. However, when the network is of small to moderate size, such consistency results are not directly applicable, which motivated various Bayesian approaches. Many of them can be viewed as variants of the latent space model \citep{hoff2002latent}, where the key idea is to assume a latent coordinate for each vertex, and the pairwise interaction of two coordinates (e.g., inner product, distance) determines the probability of whether an edge should form. Examples include the Bayesian stochastic block model [see, e.g., \cite{mcdaid2013improved,van2018bayesian,geng2019probabilistic}] that characterizes the randomness in the community labels.
Some other approaches consider edge formation as the outcome of a stochastic mechanism that can lead to a power-law degree distribution \citep{cai2016edge}, or to sparse networks \citep{caron2017sparse} and can aid in link prediction tasks \citep{williamson2016nonparametric}. 

However, in many scientific areas, it is becoming common to have access to a \emph{collection} of networks, that usually exhibit a certain degree of heterogeneity. For example, neuroscientists are collecting imaging or brain signals from EEG/MEG technologies for cohorts of patients that give rise to networks capturing brain activity between regions of interest (ROIs) \citep{shen2013groupwise}. The networks in the collection share common features (e.g., community structure, since they are derived from subjects either responding to the same stimulus in designed experimental studies or having the same disease condition in observational studies), but also exhibit heterogeneity. Analysis of such collections could proceed by
applying current approaches to each network and then devising methods for aggregating the results, which could prove challenging, since the possible significant variation from one network to another renders pooling information error-prone, for example by assuming a shared latent space. This issue was recognized by \cite{durante2017nonparametric} that proposed to use
multiple sets of coordinates, modeled by a non-parametric mixture distribution. The latter approach fits better the underlying data, vis-a-vis a naive averaging across multiple networks. Similarly, \cite{mukherjee2017clustering} proposed an approach to directly cluster the networks, which reduces the heterogeneity for downstream analysis. 

The \cite{durante2017nonparametric} work serves as our starting point, but the focus of this study is different: we are primarily interested in estimating shared community structures, providing uncertainty quantification for the estimates and also characterizing heterogeneity and its impact on the shared structures.

The focus of this work on community detection leads us to consider the graph Laplacian \citep{chung1997spectral}, which lies at the heart of spectral clustering algorithms that use a set of eigenvectors corresponding to the smallest non-trivial eigenvalues. It is a simple transformation of the adjacency matrix, and its smallest non-trivial eigenvalues provide information on the minimum edge loss when partitioning the network into multiple communities. However, spectral clustering-based approaches are primarily algorithmic in nature, involving a multi-stage procedure starting by normalizing the graph Laplacian, followed by a singular value decomposition and selection of the appropriate number of eigenvectors to use (based mostly on empirical inspection) and then finally a post-processing of the eigenvectors through an application of the K-means algorithm \citep{ng2002spectral}. Further, performance guarantees for such approaches are asymptotic in nature and take the form of high probability error bounds on the number of communities selected and the misclassification error rate \citep{hein2007graph,von2008consistency, rohe2011spectral}.  
Since there is no likelihood function involved, measures of uncertainty for community assignments are difficult to obtain, and further, it becomes challenging to accommodate heterogeneity across networks.

To address these issues, we consider a probabilistic model for the graph Laplacian, taking
advantage of its spectral properties and further introducing a non-parametric Bayes treatment on a population of networks/graphs. The crux of the problem is how to parameterize a valid Laplacian matrix
but only focusing on a small set of eigenvectors (rank) that captures the underlying community structure. We leverage ideas from the spiked covariance model \citep{donoho2018optimal}, and adding a new transformation so that we keep the few smallest eigenvalues (as opposed to the largest ones in covariance modeling). We then show that the associated eigenvectors contain useful information for a hierarchical bi-partitioning of the graph, which leads to an almost instantaneous estimation of the communities in each graph, with no need for post-processing of the results from spectral clustering with iterative algorithms like K-means. As a Bayesian model, the estimated community labels have a posterior distribution, which quantifies the uncertainty.

The remainder of this paper is organized as follows: Section 2 introduces the construction of the spiked graph Laplacian, and the non-parametric Bayesian model that accommodates the heterogeneity in a collection of graphs; Section 3 introduces the estimation of the communities based on the posterior distribution; Section 4 establishes theoretical properties for the proposed model. Section 5 evaluates the model performance based on synthetic data, while Section 6 illustrates a data application to characterize the heterogeneity in brain scans in a human working memory study.
\section{The Spiked Graph Laplacian Model}

Suppose $S$ graphs/networks are observed, each denoted by $G^{(s)}=\{V^{(s)},E^{(s)}\}, \ s=1,\ldots,S$, with corrersponding vertex set  $V^{(s)}=\{1,\ldots,n\}$ and edge set $E^{(s)}=\{e^{(s)}_{i,j}\}_{i,j}$. For ease of presentation, we focus on undirected, weighted graphs, whose adjacency matrix is given by $A^{(s)}=\{A^{(s)}_{i,j}\}_{i,j}$ with entries satisfying $A^{(s)}_{i,j}\ge 0$, $A^{(s)}_{j,i}=A^{(s)}_{i,j}$ and $A^{(s)}_{i,i}=0$. {Extension to a binary $A^{(s)}_{i,j}$ is discussed at the end of the paper.}

We start by introducing a probabilistic model for the graph Laplacian. For easing the notational burden, the graph index ${(s)}$ is omitted in the subsequent presentation. 
The observed normalized Laplacian is a transformation of the adjacency matrix given by
\bel{eq:laplacian_def}
L = D^{-1/2}(D-A)D^{-1/2},
\eel
where $D=\diag{d_{i}}_{i=1}^n$ is the observed degree matrix, with $d_i=\sum_{i=1}^n A_{i,j}$. Viewing $L$ as a noisy observation, we consider the following signal-plus-noise matrix model:
\bel{eq:spiked_laplacian}
L  = \mu_L + \mathcal E, \qquad & \mu_L=\sum_{k=1}^T \lambda_k  \vec q_k  \vec  q_k^{\prime}
        + \sum_{l=T+1}^n \theta  \vec q_l \vec q_l^{\prime} ,
\eel
where $\mathcal E$ is a symmetric matrix capturing random variation with $\mathcal E=\{e_{i,j}\}_{i,j}, e_{i,j}\sim \No(0,\sigmae)$ for $i<j$. The matrix $\mu_L$ is symmetric with eigenvalues $\lambda_1, \lambda_2  \ldots,\lambda_T, \underbrace{\theta,\ldots,\theta}_{(n-T)}$ and corresponding eigenvectors $\vec q_1,\ldots,\vec q_n$.

We consider $\mu_L$ to be the graph Laplacian of an unobserved ``true" graph $A_*$ without measurement noise:
$
\mu_L = D_*^{-1/2}(D_*-A_*)D_*^{-1/2},
$
with $D_*$ being its degree matrix.

\begin{lemma}[The first eigenvector of the Laplacian]
 The normalized Laplacian $\mu_L= D_*^{-1/2}(D_*-A_*)D_*^{-1/2}$ has all eigenvalues $\lambda_k\in[0,2]$ and 
$\theta\in[0,2]$. The smallest eigenvalue $\lambda_1=0$ (index 1 denotes the smallest one) and its corresponding eigenvector 
$\vec q_1=\vec d^{1/2}_*= (d_{*1}^{1/2},\ldots, d^{1/2}_{*n})^{\prime}$. That is,
      $$
      \mu_L \vec d_*^{1/2}= 0 \cdot \vec d_*^{1/2}.
      $$
\end{lemma}
Therefore, the Laplacian  $\mu_L$ is a sufficient statistic of $A_*$, as we can recover
\bel{eq:true_graph}
D_* = \diag{q^2_1(i)}_{i=1}^n, \quad A_* = D_*^{1/2} (I- \mu_L) D_*^{1/2},
\eel
where $q_1(i)$ denotes the $i$th element in $\vec q_1$. Similarly, we can recover $A$ deterministically using the observed $L$ matrix and its first eigenvector $\vec {\hat q}^2_1$
\bel{eq:observed_graph}
D = \diag{\hat q^2_1(i)}_{i=1}^n, \quad A = D^{1/2} (I- L) D^{1/2}.
\eel
Therefore, the probabilistic model \eqref{eq:spiked_laplacian} for $L$ is also the complete model for the graph adjacency matrix. If imposing additional constraint on $\mathcal E$, such that $L_{i,j}\le 0$ for all $i< j$, then we could use \eqref{eq:spiked_laplacian} and \eqref{eq:observed_graph} as a generative graph. Since in this paper, we assume $L$ is observed and fixed, we choose to ignore this constraint for model simplicity.

To satisfy $d_{*}^{1/2}(i)>0$, we now further require $q_1(i)>0$ for all $i$. Collecting the eigenvectors and eigenvalues 
in matrices $Q=(\vec q_1,\ldots, \vec q_T)$ and $\Lambda= \text{diag}(\lambda_1,\ldots,\lambda_T)$, and after re-arranging terms in \eqref{eq:spiked_laplacian} we obtain
\bel{eq:spiked_laplacian_matrix}
\mu_L &= \sum_{k=1}^T (\lambda_k -\theta )\vec  q_k  \vec q_k^{\prime}
        + \sum_{l=1}^n \theta   \vec q_l \vec q_l^{\prime} \\
        & = Q(\Lambda -I_T\theta)Q^{\prime} + I_n\theta,
\eel
where $\vec q_{T+1},\ldots, \vec q_n$ are canceled due to orthonormality, $\sum_{l=1}^n  \vec q_l \vec q_l^{\prime}=I_n$; therefore, this leads to a substantial reduction in the number of model parameters from $O(n^2)$ to $O(nT)$.

\begin{remark}
This model is largely inspired by the spiked covariance model \citep{donoho2018optimal}, except that the ``spikes'' $\lambda_2,\ldots, \lambda_T$ are associated with the \emph{smallest} eigenvalues, that as shown later, drive the partitioning of the graph into communities. For this reason, we coin the term ``spiked graph Laplacian'' for $\mu_L$.
\end{remark}

\subsection{A Non-parametric Bayesian model for Heterogeneous Spiked Graph Laplacians}

A key benefit of the probabilistic model introduced for the spiked graph Laplacian is that it enables us to
naturally capture heterogeneity in a collection of such graphs. We consider a collection of graphs $G^{(s)}, s=1,\cdots,S$,
with associated Laplacians and their decompositions $(\mu_L^{(s)},Q^{(s)},\theta^{(s)},\Lambda^{(s)})$.

Note that in \eqref{eq:spiked_laplacian}, each $\vec q_k^{(s)}$ forms a factor matrix 
$\vec  q_k^{(s)}  \vec q_k^{(s)\prime} $ encoding the pairwise {interactions of the vertices}, while each $\lambda^{(s)}_k$ 
modulates the magnitude of the {interactions}. 

Given such a heterogeneous collection of graphs, in order to not only learn the common community structure across them but also
capture their heterogeneity as reflected in their edge density, we use a two-fold approach; specifically, a
non-parametric Bayes model is used for estimating a common dictionary of factors, while a random-effects model
controls the number of spikes for each graph Laplacian $L^{(s)}$.

The matrix of eigenvectors $Q^{(s)}$ is modeled based on a Dirichlet process mixture, 
\bel{eq:eg_matrix}
&Q^{(s)} \sim \sum_{l=1}^\infty \pi_l \delta_{U^{(l)}} (.), \qquad \Pi(U^{(l)} ) \propto  \exp  \big\{ \tr \big[   \Omega M ^{\prime}U^{(l)} \big] \big\}  \mathbf{I}[ u^{(l)}_1(i)>0 \text{ for } i=1,\ldots,n], \\
& \pi_1 = \nu_1, \quad \pi_l = \nu_l \prod_{l'<l} (1-\nu_{l'}) \text{ for } l>1, \\
& \nu_l \sim \text{Beta}(1,\alpha_0),
\eel
which has the base measure from a constrained matrix Langevin distribution, on a Stiefel sub-manifold with the elements in first column being all positive, $\mathcal{V}^{T, n}_*=\{Q\in\mathbb{R}^{n\times T}:Q'Q=I_T, q_1(i)>0,i=1\ldots n\}$; $\Omega$ a diagonal $T\times T$ matrix; the concentration parameter $\alpha_0>0$; $\delta_a(.)$ a point mass  at $a$; $\mathbf{I}(E)$ takes value $1$ if $E$ holds, otherwise taking $0$.

An important property of the Dirichlet process mixture is that the posterior distribution is discrete almost surely \citep{sethuraman1994constructive}.
Therefore, using this non-parametric Bayes prior allows us to obtain a discrete distribution for $Q^{(s)}$, where $Q^{(1)} \ldots, Q^{(S)}$ have only a few unique values much less than $S$. That is, we learn a ``dictionary'' of the eigenmatrices.

The eigenvalues $\lambda^{(s)}_k $ and $\theta^{(s)},  s =1,\dots,S $ are assumed independently and identically distributed according to the following prior distribution:
\bel{eq:eg_value}
&\eta_k^{(s)}\sim \text{Bernoulli}(w),\\
& \lambda^{(s)}_k  \mid \eta_k^{(s)}=1 \sim \text{No}_{(0,2)} (0,\sigma^2_{\lambda,1}) ,
\quad \lambda^{(s)}_k  \mid \eta_k^{(s)}=0 \sim \text{No}_{(0,2)} (\mu_\theta,\sigma^2_{\lambda,0}), \\
&\theta^{(s)} \sim \No_{(0,2)}(\mu_\theta, \sigma^2_\theta),
\eel
for $ k=2,\ldots, T$, with $\No_{(0,2)}$ denoting a Gaussian distribution truncated to the $(0,2)$ interval. Further,
since $\lambda_1^{(s)}=0$, we assign $ \eta^{(s)}_1=1$. When marginalizing over $\eta_k^{(s)}$,
each $\lambda^{(s)}_k$ follows a two-component mixture, with the first component capturing small spikes
close to 0, and the second component spikes close to $\theta$. This enables  a constant dimension $T$ for all $L^{(s)}$, while retaining adaptiveness to have the effective number of small spikes:
\bel{eq:effective_com}
\kappa^{(s)} = \sum_{k=1}^{T} \eta^{(s)}_k,
\eel
as shown later, equivalent to $\kappa^{(s)}$ communities. 

\begin{remark}
An alternative parameterization
would be using $T^{(s)}$ that varies directly with each graph; however, this would lead to an inefficient discrete search when estimating
the posterior distribution.
\end{remark}

\begin{figure}[H]
 \begin{subfigure}[t]{0.22\textwidth}
 \centering
       \caption*{\small (a) }
       \includegraphics[width=1\linewidth, height=3cm]{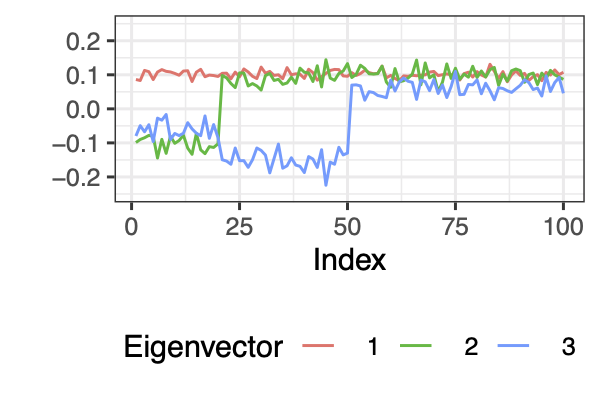}
 \end{subfigure}
 \begin{subfigure}[t]{0.22\textwidth}
 \centering
        \caption*{\small (b) }
       \includegraphics[width=1\linewidth, height=3cm]{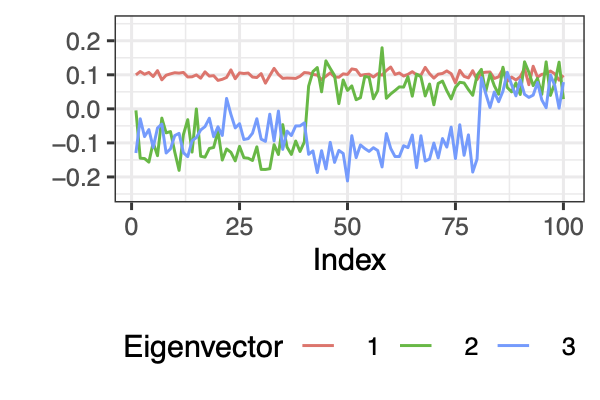}
 \end{subfigure}
\begin{subfigure}[t]{0.22\textwidth}
 \centering
         \caption*{\small (c) }
       \includegraphics[width=1\linewidth, height=3cm]{sim1_u2.png}
 \end{subfigure}
 \begin{subfigure}[t]{0.22\textwidth}
 \centering
          \caption*{\small (d) }
       \includegraphics[width=1\linewidth, height=3cm]{sim1_u2.png}
 \end{subfigure}
 \vspace*{-5mm}
\caption*{\footnotesize 4 eigenmatrices $Q^{(s)}$ sampled from \eqref{eq:eg_matrix} }
 \begin{subfigure}[t]{0.22\textwidth}
 \centering
       \includegraphics[width=1\linewidth]{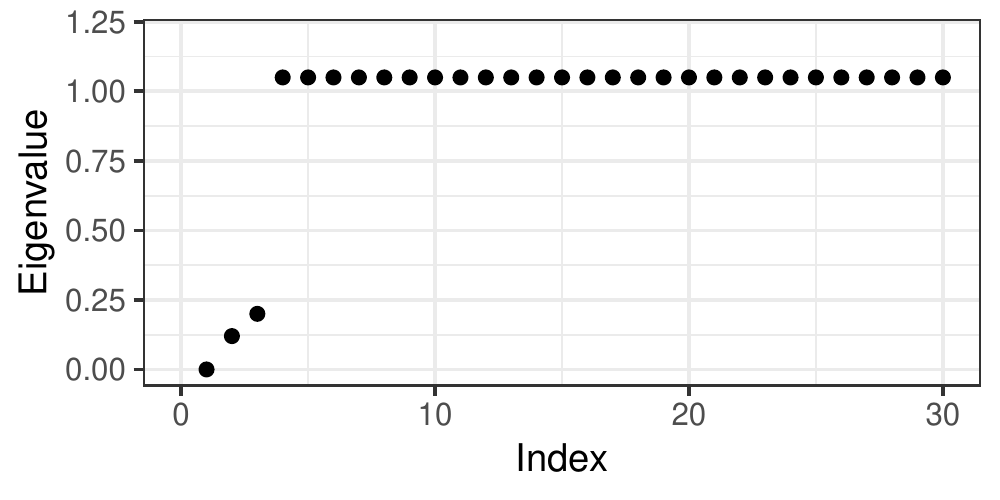}
 \end{subfigure}
 \begin{subfigure}[t]{0.22\textwidth}
 \centering
       \includegraphics[width=1\linewidth]{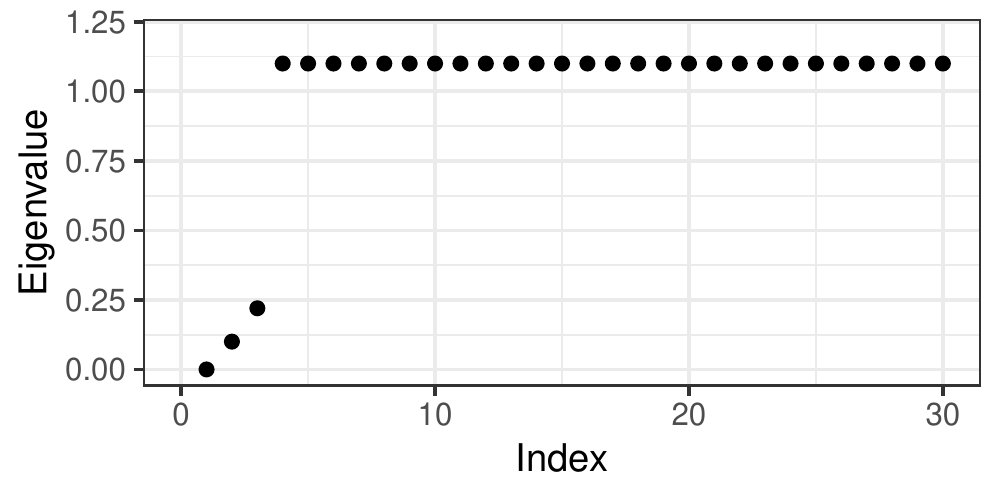}
 \end{subfigure}
\begin{subfigure}[t]{0.22\textwidth}
 \centering
       \includegraphics[width=1\linewidth]{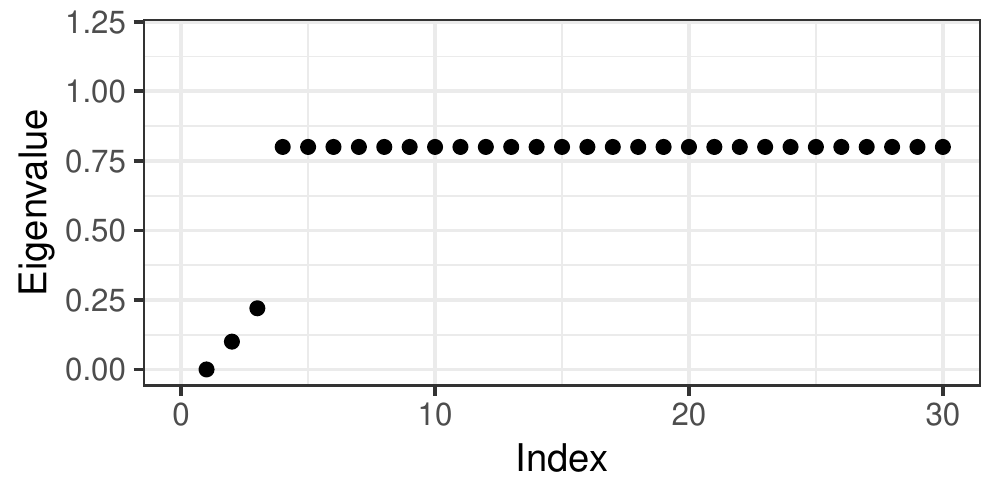}
 \end{subfigure}
 \begin{subfigure}[t]{0.22\textwidth}
 \centering
       \includegraphics[width=1\linewidth]{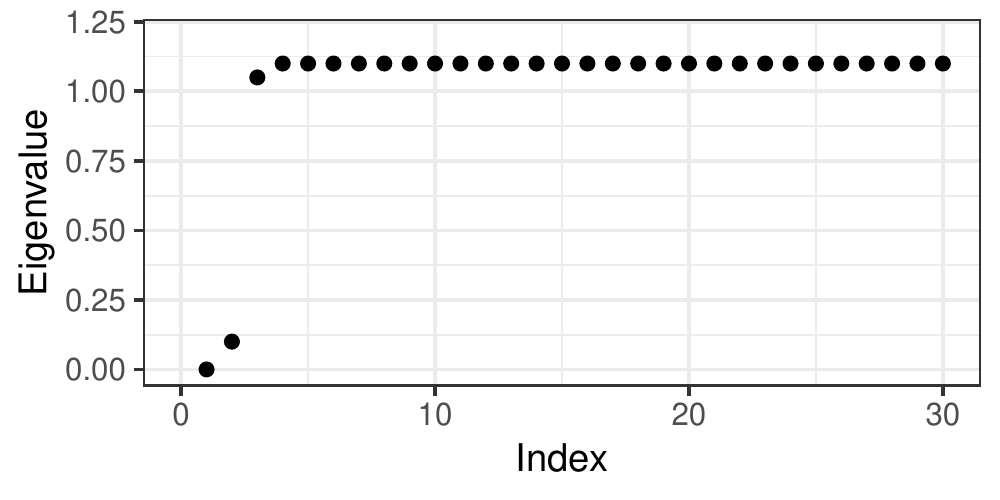}
 \end{subfigure}
  \vspace*{-5mm}
 \caption*{\small 4 sets of eigenvalues $(\Lambda^{(s)},\theta^{(s)})$ sampled  from \eqref{eq:eg_value} .}
  \vspace*{-5mm}
 \begin{subfigure}[t]{0.22\textwidth}
 \centering
       \includegraphics[width=1\linewidth]{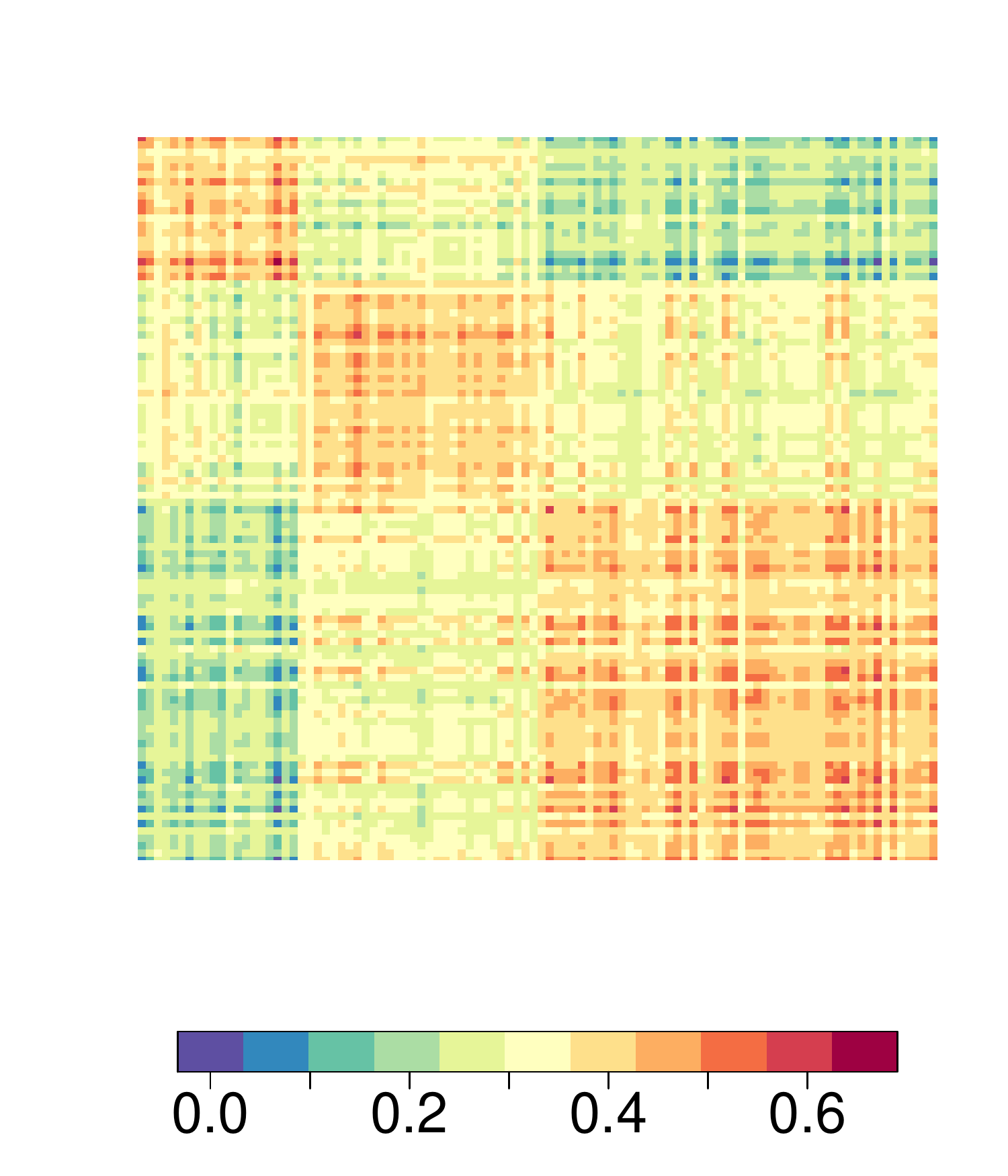}
 \end{subfigure}
  \,
  \begin{subfigure}[t]{0.22\textwidth}
 \centering
       \includegraphics[width=1\linewidth]{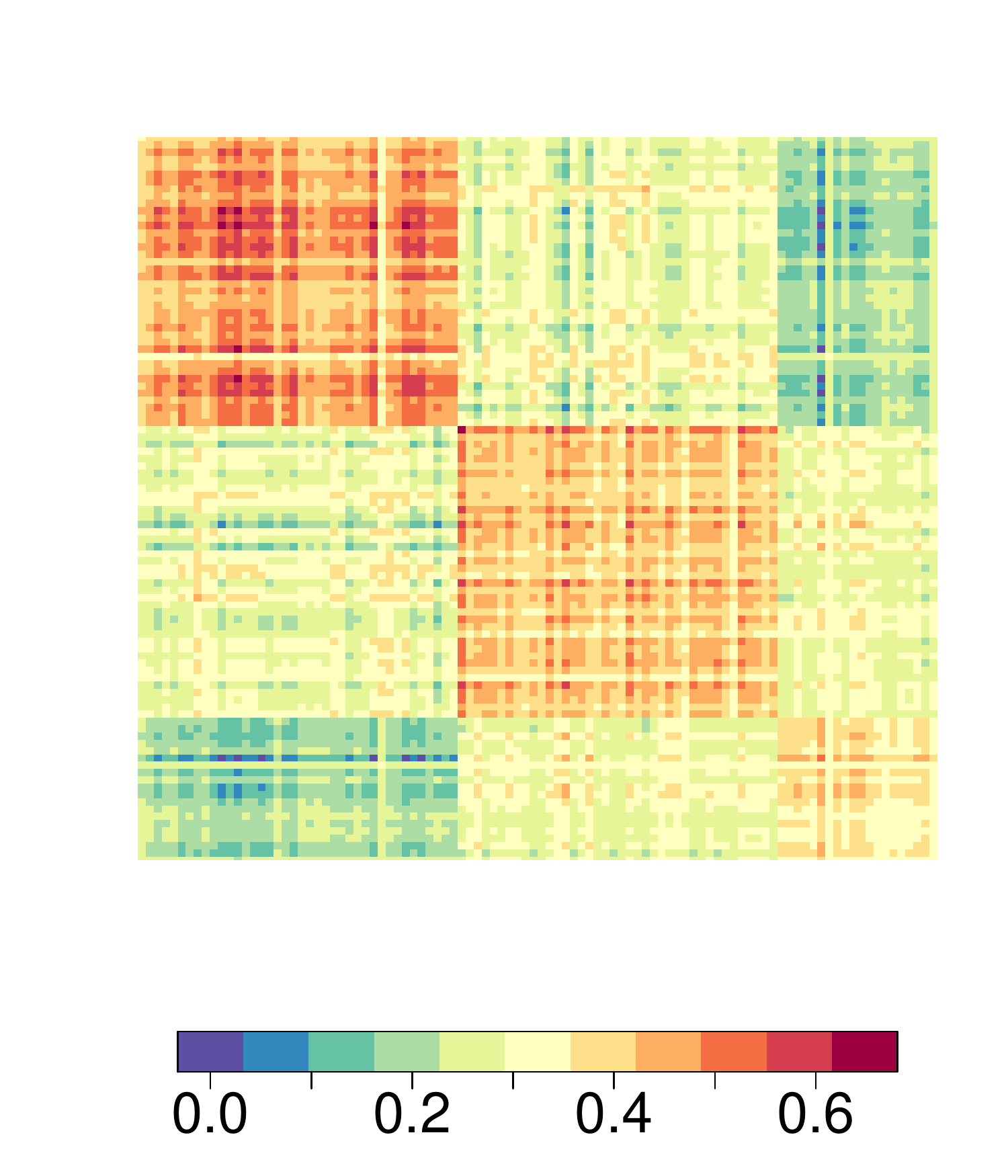}
 \end{subfigure}
  \;
\begin{subfigure}[t]{0.22\textwidth}
 \centering
       \includegraphics[width=1\linewidth]{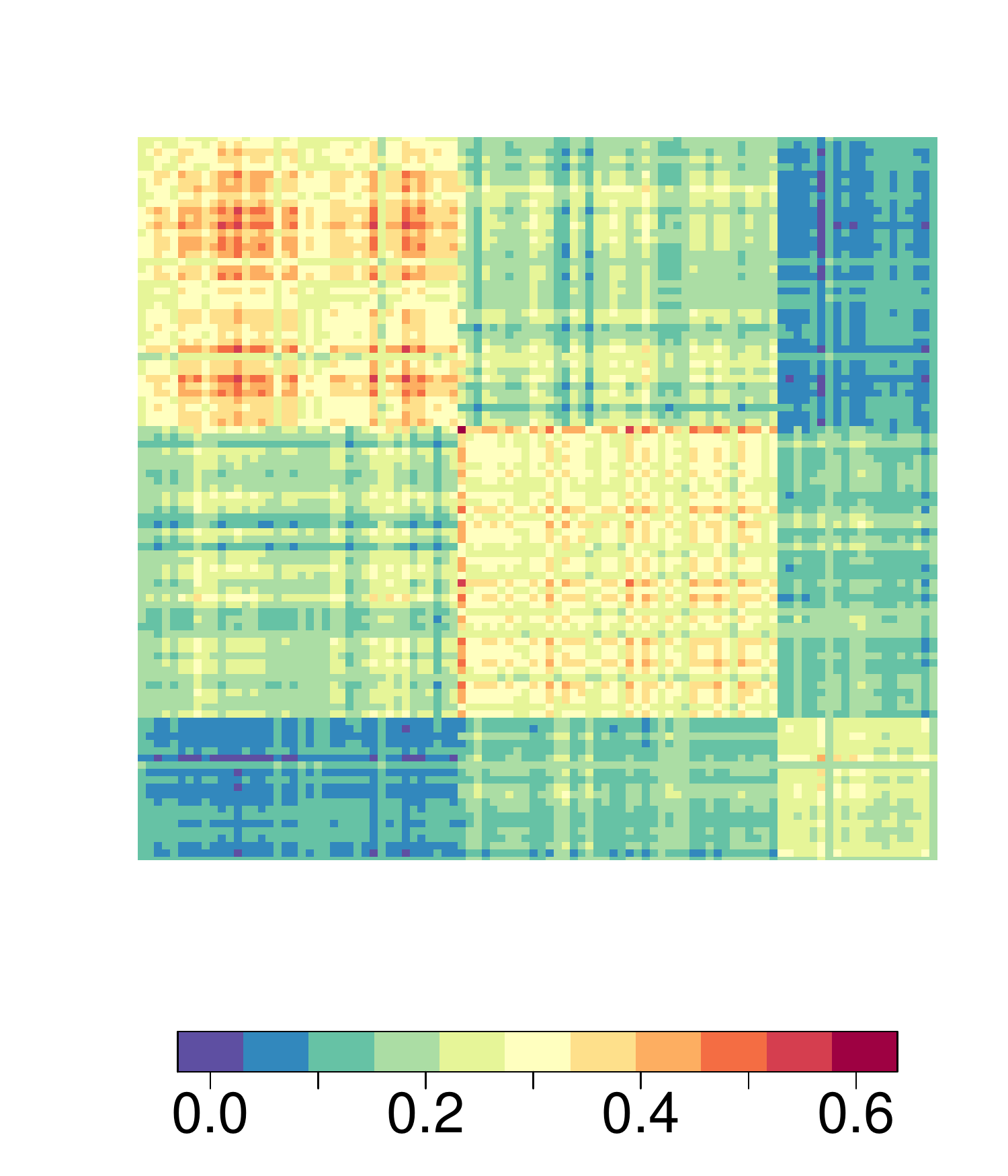}
 \end{subfigure}
 \;
 \begin{subfigure}[t]{0.22\textwidth}
 \centering
       \includegraphics[width=1\linewidth]{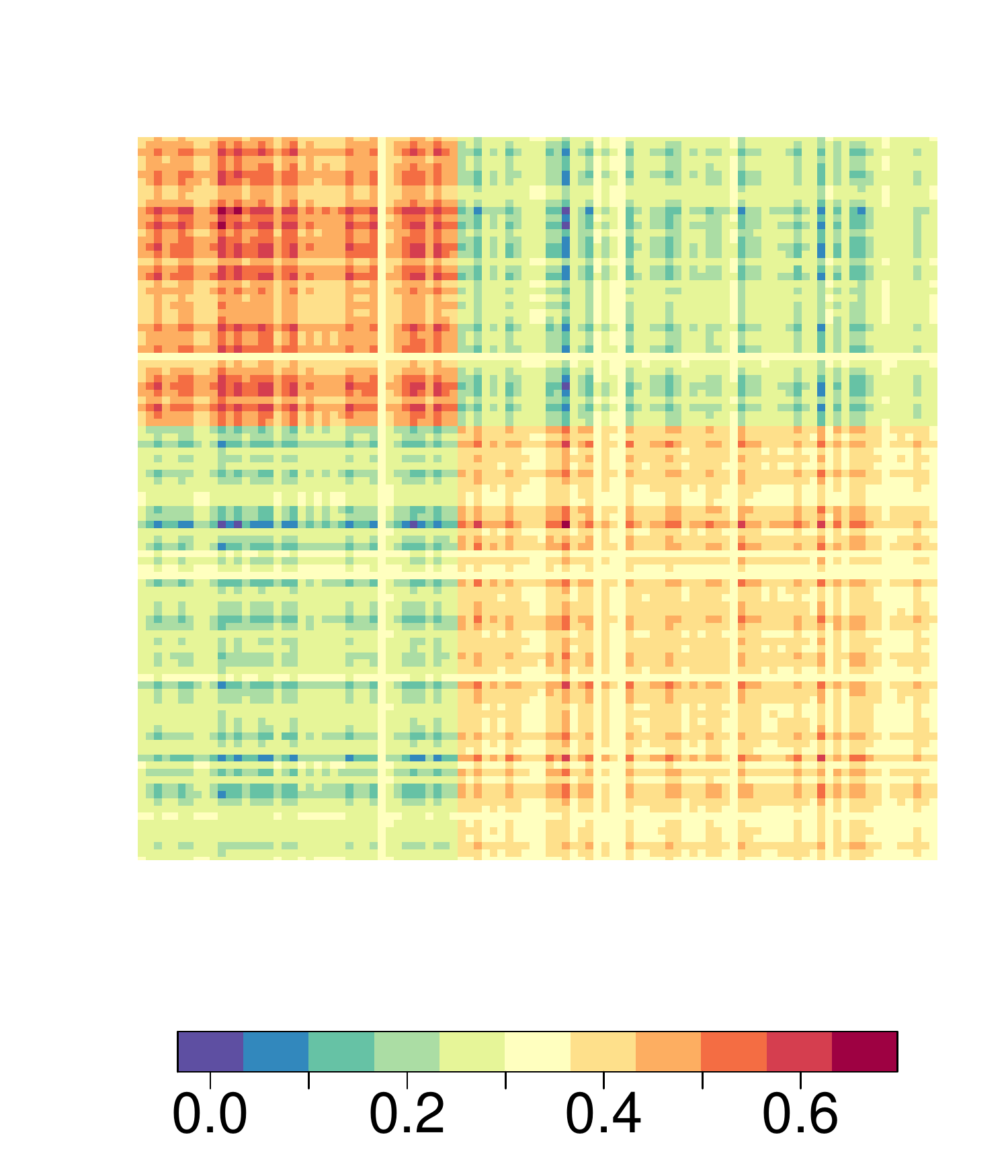}
 \end{subfigure}
  \vspace*{-5mm}
 \caption*{\small Corresponding adjacency matrices simulated using $L  = \mu_L + \mathcal E$, $A=D^{1/2}(I-L)D^{1/2}$.}
  \vspace*{-1mm}
  \caption{\footnotesize Simulation illustrating how the non-parametric spiked Laplacian model addresses graph heterogeneity: Graphs (b), (c), (d) use the same eigenmatrix $Q^{(s)}$ drawn from the Dirichlet process, creating similar community structure; the independent eigenvalues $\Lambda^{(s)}$ lead to varying degree of sparsity ((b) vs (c)) and also dictate whether a community can be further divided into two smaller communities ((b) vs (d)). Graph a take a different value for $Q^{(s)}$; hence its community structure is completely different from (b), (c), (d).  \label{fig:sim1}}
 \end{figure}

Next, we illustrate the high flexibility of the proposed modeling framework
based on synthetic data. We draw four eigenmatrices
from \eqref{eq:eg_matrix} and four sets of eigenvalues from \eqref{eq:eg_value},
and obtain the adjacency matrix using \eqref{eq:spiked_laplacian} and \eqref{eq:observed_graph}
with $\sigma^2_e =10^{-2}$. As shown in Figure~\ref{fig:sim1}: (i) Graphs
(b), (c), (d) have the same values in the eigenmatrix $Q^{(s)}$, therefore
they share a similar community structure and appeare quite different from graph (a);
(ii) among those three, the independent eigenvalues $\Lambda^{(s)}$ create
 varying edge weights thus leading to different strengths in connectivity between graphs (b) and (c), and also dictate whether a community can be further
divided into two smaller communities [(b) vs (c)].

\subsection{Specification of the Prior Distribution}
For the variance parameters $ \sigma^2_\theta$, $\sigma^2_{\lambda,0}$ and $\sigma^2_{\lambda,1}$, we assign proper $\text{Inverse-Gamma}(2,
 0.1)$ with a weakly informative prior mean at $0.1$. For the mean parameter $\mu_\theta$, we set it to $1$, to ensure identifiability, so that $\theta$ remains bounded away from $0$ and also reflects the prior belief that it is around the center of the interval $[0,2]$ that the eigenvalues of the Laplacian take values in. 
 For $w$, we assign a non-informative  prior $\text{Beta}(1,1)$. For the  noise variance
$\sigma^2_e$, we set a diffuse prior $\text{Inverse-Gamma}(0.01,0.01)$. For the base measure of the Dirichlet process  \eqref{eq:eg_matrix}, we choose the non-informative  $\Omega =\text{diag}(0,\ldots,0)$,  making it a uniform prior
measure over $\mathcal{V}^{T, n}_*$ and eliminating the need to estimate $M$ or any intractable normalizing constant.  We choose concentration $\alpha_0=0.1$ to induce sparsity in the mixture weights, which lead to fewer unique values in $Q^{(s)}$ thus aiding interpretation.

In numerical experiments, this prior specification shows good empirical performance in recovering the ground truth and is robust to a wide range of values of $n$, $S$ and noise levels without the need for tuning.  
\subsection{Estimation of the Posterior Distribution}
We use Gibbs sampling to estimate the posterior distribution. 
Since an infinite mixture distribution is involved, we use a latent assignment $z_s \in \{1,2, \ldots \}$ for each graph, such that $Q^{(s)}=U^{(l)}$
if $z_s=l$. Then, the likelihood given $\{z_s\}$ becomes
\bel{eq:conditional_likelihood}
\prod_{s=1}^S\Pi(L^{(s)}; \quad &\sigma_e^2, \Lambda^{(s)},Q^{(s)} , \theta^{(s)}, z^{(s)})  \\
& \propto 
 (\sigma_e^2)^ {- \frac{S n(n+1)}{4}}
\exp\bigg( - \sum_{s=1}^S \frac{1}{4\sigma_e^2}
\bigg \{ \tr\big[
  (\Lambda^{(s)} -\theta^{(s)} I_T)^2 \big] + \|L - \theta^{(s)}  I_n \|_F^2  \bigg\}\\
 & + \sum_{l=0}^\infty \sum_{s: z_s =l}  \frac{1}{2\sigma_e^2}\text{tr}
\big[( \theta^{(s)}  I_n - L^{(s)}) U^{(l)} (\theta^{(s)}I_T- \Lambda^{(s)} ) U^{{(l)}\prime}\big] \bigg).
\eel

In the above, we replaced the fixed diagonal elements $L^{(s)}_{i,i}= 1$ with an augmented random variant $L^{(s)}_{i,i}=\No(\mu_{L,i,i} , 2 \sigma_e^2)$, for easier matrix-based computation as in \cite{Hoff:2009eg}.

\emph{A Gaussian Integral Trick for the Product-Matrix-Bingham Distribution:}
One immediate challenge of sampling $U^{(l)}$ from \eqref{eq:conditional_likelihood} is the exponential-quadratic in the full conditional: 
\bel{eq:posterior_U}
\Pi(U^{(l)} \mid .) \propto \exp\bigg\{ \frac{1}{2\sigma_e^2} \sum_{s: z_s
=l} \tr( F_s U^{(l)} G_s U^{(l)\prime})\bigg\} \etr(\Omega M^{\prime} U^{(l)}),
\eel
 where $F_s = \theta^{(s)} I_n - L^{(s)} $ and $G_s =
\theta^{(s)}I_T- \Lambda^{(s)} $. This corresponds to 
the product of a matrix Bingham-$\{F_s/(2\sigma_e^2), G_s\}$,
which lacks a closed form for sampling.

To solve this problem, we propose a
new data augmentation for the product-matrix-Bingham distribution, which extends the Gaussian integral trick  \citep{zhang2012continuous} on the Stiefel manifold. Consider an augmented random matrix $R_s\in \mathbb{R}^{T\times n}$ from the matrix Gaussian $\text{Mat-No}(G_sU^{\prime}F_s, G_s \sigmae, F_s)$:
\bel{eq:aug_matrix_gaussian}
\Pi( R_s \mid U^{(l)}) \propto |F_s|^{-T/2} |G_s|^{-n/2}\etr \bigg\{  -\frac{1}{2\sigmae} F_s^{-1} (R_s-G_sU^{(l)\prime}  F_s )^{\prime} G_s^{-1} (R_s-G_s U^{(l)\prime} F_s ) \bigg\},
\eel
Given $R_s$, all quadratic terms in \eqref{eq:posterior_U} are canceled, leading to
\bel{eq:aug_posterior}
\Pi(U^{(l)} \mid \{R_s\}_{s:z_s=l}) \propto  \etr\bigg( \frac{1}{\sigmae}  \sum_{s:z_s=l}   R_s U ^{(l)}+ \Omega M^{\prime}  U ^{(l)}\bigg).
\eel
Therefore, we can sample \eqref{eq:aug_matrix_gaussian} and \eqref{eq:aug_posterior} alternatively in closed form; and the latter is a matrix Langevin distribution amenable to the sampling algorithm in \cite{Hoff:2009eg}.

\emph{Sampling Algorithm:} To simplify computations, we approximate the Dirichlet process mixture
model with a truncated version, setting the number of mixture components to $g$ and using $\text{Dir}(\alpha_0/g,\ldots, \alpha_0/g)$ (in this paper, we use $g=30$).
The detailed steps of the algorithm are given in the supplementary materials.

\vspace{-10mm}
\section{Community Detection based on the Posterior Distribution}
In this section, we focus on the community assignment labels $c_i^{(s)}\in \mathbb{N}$ for each vertex $i$ in graph $s$, using the obtained posterior sample of $Q^{(s)}$ and $\Lambda^{(s)}$. Specifically, we obtain $\{ c_i^{(s)} \}_{i=1}^n$ via a fast and deterministic transformation of $(Q^{(s)}, \Lambda^{(s)})$, in which the algorithm aims to optimize the partitioning of each graph; in the meantime, since this is a measurable transformation, we quantify the uncertainty via the induced  distribution of $c_i^{(s)}$. Since the discussion pertains to each graph, we omit superscript $(s)$ for ease of presentation.

\noindent
\emph{Optimal Graph Cut}

We first introduce the concept of ``optimal graph cut"'. In the simplest possible case, suppose we want to bi-partition (or,   ``cut") a graph $G=(V, E)$ into two sub-graphs $G_1 = \big (V_1, E(V_1,V_1) \big)$ and $G_2 = \big(V_2, E(V_2, V_2)\big)$, with $V_1\cup V_2=V$ and $V_1\cap V_2=\varnothing$, and $E(V_j,V_j)$ the edges formed among the vertices within $V_j$.
 An intuitive cut criterion corresponds to minimizing the loss of edge weights between two sub-graphs: $\sum_{i \in V_1, j \in V_2} A_{i,j}$.

On the other hand, we want to prevent trivial cuts, where one of the partition vertex sets $V_j, j=1,2$ comprises of few or even a single vertex.
To that end, \cite{shi2000normalized} introduced the minimal \emph{normalized cut} loss defined as
\be
h_2(G) = \underset{(V_1,V_2)}{\min}\frac{\sum_{i \in V_1, j \in V_2} A_{i,j}}{\min_{l=1,2} \sum_{ i,j \in V_{l} } A_{i,j}},
\ee
where the denominator is the sum of the vertex degrees in one of two subgraphs. Initially, $h(G)$ was proposed for a binary adjacency matrix $A$, and is also known as the Cheeger or isoperimetric constant \citep{mohar1989isoperimetric},       
representing the bottleneck of the flow across the edges connecting the two partitioned vertex sets; later on, this loss was extended to weighted graphs \citep{friedland2002cheeger}.

 \cite{louis2011algorithmic} extends it to $\kappa$-partitioning of a weighted graph, with the corresponding loss function known as the ``sparsest $\kappa$-cut'':
\be
h_\kappa(G) = \min_{(V_1,\ldots, V_\kappa)}\frac{\sum_{m<l}\sum_{i \in V_m, j \in V_l} A_{i,j}}{ \min_{l=1,\ldots,\kappa} \sum_{ i,j \in V \setminus V_{l} } A_{i,j}},
\ee
where $(V_1,\ldots, V_\kappa)$ is a partitioning of $V$.

Interestingly, the optimal values of these losses are upper-bounded by the eigenvalues of the graph Laplacian. Consider the graph associated with the adjacency $A^*$ in \eqref{eq:true_graph}; we then have
\bel{eq:cheeger_bound}
h_2(G)\le \sqrt{2{\lambda_{(2)}}}, \qquad h_\kappa(G)  \le (8\log \kappa ) \sqrt{\lambda_{(\kappa)}} \text{ for } \kappa\ge 3,
\eel
where the former is due to \cite{friedland2002cheeger}, and the latter due to  \cite{louis2011algorithmic}, with 
$\lambda_{(k)}$ denoting the $k$-th smallest eigenvalue in $\{\lambda_1,\ldots \lambda_T\}$.

Recall that in the spiked graph Laplacian model, there are $\kappa$ small spikes; see, \eqref{eq:effective_com}.
Hence, since $\lambda_{(1)},\ldots,\lambda_{(\kappa )}\approx 0$, $\kappa$ communities can be extracted 
with negligible graph-cut loss.

\subsection{Sign-based Partitioning}

Finding the best $\kappa$-partition is a challenging problem computationally, due to the combinatorial search required.
However, there are numerous algorithms in the literature that approximate the optimal cut. Examples include the spectral clustering \citep{ng2002spectral} and the random search algorithm \citep{louis2011algorithmic}. In particular, the latter one is shown to achieve a loss smaller than $(8\log \kappa
) \sqrt{\lambda_{(\kappa)}}$  for any ${\lambda_{(\kappa)}}$, although the computations involved can be intensive.

\begin{algorithm}[H]
\caption{Sign-based $\kappa$-partitioning.\label{algo}}
\begin{algorithmic}[]
\STATE{Initialize: $V_{[1]1}= \{1,\ldots,n\}$, re-order $\{\vec q_k\}_{k=1}^T$ according to ascending order of $\lambda_k$, denoted by $\{\vec q_{(k)}\}_{k=1}^T$}.
   \FOR{$k=1$ \TO $(\kappa-1)$ } 
   \STATE {1. Compute the loss for $l=1,\ldots,k$
     \be loss_{[k]l}= \sum_{i,j \in V_{[k]l} }  \big[q_{(k)}(i)q_{(k)}(j)\big] 1\big [q_{(k)}(i) q_{(k)}(j) < 0\big].
     \ee
   }
   \STATE{2. 
   Find $l^*= \underset{l\in \{1\ldots, k\}} {\arg\min} \, loss_{[k]l}$, add one partition by setting 
  \be   
&  V_{[k+1]l^*} :=\{i\in V_{[k]l^*} q_k (i)\ge 0\} \\
&  V_{[k+1](k+1)}  :=\{i\in V_{[k]l^*}: q_k (i)< 0\}\\
&  V_{[k+1]l} :=V_{[k]l} \text{ for } l\neq l^*, l\le k.
  \ee
   }
    \ENDFOR
   \STATE{Use  $\{V_{[\kappa]l}\}_{l=1}^{\kappa}$ as the $\kappa$-partition; record $c_i=l$ for $i\in V_{[\kappa]l}$.}
\end{algorithmic}
\end{algorithm}
Inspired by the famous Fiedler vector
\citep{fiedler1989laplacian},
we propose a more efficient algorithm using the signs in the eigenvectors (see Algorithm~\ref{algo}). 

The justification for the key steps in the proposed algorithm is as follows. Examine the off-diagonal elements of each adjacency matrix \eqref{eq:true_graph}
\bel{eq:vec_loss}
A_{*,i,j}=d_i d_j\sum_{k=1}^T
(\theta-\lambda_{(k)}) q_{(k)}(i) q_{(k)}(j) , \qquad
i\neq j,
\eel
for $d_i>0, d_j>0$ and $(\theta-\lambda_{(k)}) >0$ for small $\lambda_{(k)}$. If $q_k(i)$ and $q_k(j)$ have the same sign, they contribute positively to $A_{*,i,j}$. Therefore, to minimize the loss due to a graph cut, a locally optimal cut is simply dividing the set into two subsets --- the one with $q_{(k)}\ge 0$ and the one with $q_{(k)}<0$. In the simplest case with $\kappa=2$, this is exactly the Fiedler vector partitioning \citep{fiedler1989laplacian}. We do this recursively until obtaining $\kappa$ subsets.

Due to the orthonormality of the eigenvectors, the following holds for $k\ge 2$,
\be
\|\vec q_{(k)}\| =1, \qquad   \sum_{i=1}^n q_{(1)}(i)q_{(k)}(i) =0, \qquad q_{(1)}(i)>0.
\ee
To satisfy these constraints, each vector $q_{(k)}$ must contain both plus and minus signs; hence, we can always use the sign-based partitioning. This algorithm can run very fast, since it only takes one scan from $1$ to $\kappa$.
\section{Theoretical Results}

Next, we establish several properties of the proposed methodology. We first show that adapting $\kappa^{(s)}$ for the graph involves a trade-off between the number of eigenvectors to be estimate and their estimation accuracy compared to the ground truth under noise perturbations; hence, this is a trade-off between the number of communities $\kappa^{(s)}$ one attempts to identify and the uncertainty/error on the estimated community membership $c^{(s)}_i$.

Assume $L$ is a noisy version of an underlying oracle $L_0$ (not necessarily having a spiked structure), with $Q_0$ containing its eigenvectors. The spiked graph Laplacian model produces an posterior estimate 
$\hat L = \hat Q (\lambda - I_T \theta) \hat Q^{\prime} + I_n \theta$, equipped with a posterior distribution.  We can quantify the distance between the sub-matrices of  $\hat Q$ and  $Q_0$.
\begin{theorem}
[Trade-off between resolution and estimating accuracy]
For any given posterior sample from the spiked graph Laplacian model, 
let the eigen-vectors/values in the spiked graph Laplacian estimate
be ordered such that $\lambda_1\le \lambda_{2} \le \lambda_{3}\ldots \le \lambda_{T} < \lambda_{T+1}=\ldots=\lambda_{n} =\theta$. Furtherm assume each element of $(\hat L - L_0)$ is $\sigma_{e}$-sub-Gaussian, due to both $L_0$ and $\hat L$ being normalized Laplacians and thus all their elements are in the  $[-1,0]$ interval. Denote the sub-matrices formed by the first $k$ columns of the $\hat Q$ and $Q_0$ matrices as $\hat Q_{1:k}$ and $ Q_{0,1:k}$, respectively;
then for any  $k\in [2,T-1]$, there exists an orthonormal matrix $O$
    \be
 \textup{Pr}  \bigg( \| \hat Q_{1:k}O  - Q_{0,1:k}  \|_F \le \frac{\sqrt{k n}2^{3/2}  \sigma_{e}}{ \lambda_{k+1} - \lambda_{k}}  t \bigg) \ge 1-\delta_t,
    \ee
 where $\delta_t = \exp[ -  \{t^2/64 - \log(5 \sqrt{2})\} n ]$.
\end{theorem}
\begin{remark}
  The theorem shows that the following two factors are important: (i) a sufficiently large $T$ that produces a good fit, and thus a small $\sigma_{e}$; (ii) a choice for $k$ not too large, but also satisfying $\lambda_{(k+1)} - \lambda_{(k)}$ bounded away from zero. In our model \eqref{eq:eg_value}, this coincides with our choice of $k:=\kappa$ such that $\lambda_\kappa\approx 0$ and   $\lambda_{\kappa+1}\approx \mu_\theta$.
\end{remark}

Next, note that the likelihood function can be re-written as,
\be
\Pi(L; \sigma_e^2, \Lambda,Q , \theta) \propto 
 (\sigma_e^2)^ {- n(n+1)/4}
\exp \bigg( 
- \frac{1}{4\sigma_e^2}\big [
\tr (\Lambda\Lambda) - 2 \tr(\Lambda Q'LQ)
\big]
\bigg)
\\
\exp \bigg( 
- \frac{1}{4\sigma_e^2}\big [
   \theta^2(n-T)
- 2\theta \tr\{ L(I-QQ')\}
\big]
\bigg)\exp\big\{ - \frac{1}{4\sigma_e^2}\tr(LL)\big\},
\ee
where $\Lambda$ and $\theta$ are conditionally independent. Integrating out $\Lambda$ and $\theta$, we obtain the marginal likelihood.

\begin{theorem}
The marginal likelihood of $Q$ is given by
 \be
 & \Pi(L; \sigma_e^2,Q) \propto \exp 
 \bigg\{
 \frac{(\sum_{k=T+1}^n
q_k^{\prime}
L
q_k
 )^2  }{4\sigmae (n-T)}   
\bigg \}
  \exp\bigg\{ 
 \frac{  \sum_{k=1}^T(q_k^{\prime}L  q_k)^2}{4\sigmae}
 \bigg\} \zeta,
 \ee
 with $\Phi$ denoting the cumulative distribution function of the normal distribution and 
 \be
\zeta=&(\sigmae)^{- {n(n+1)/4+(T+1)/2}}
    \prod_{k=2}^T  \{\Phi( \frac{2- q_k^{\prime}L  q_k}{\sqrt{{2\sigma^2_e}} }) -
  \Phi(\frac{- q_k^{\prime}L  q_k}{\sqrt{{2\sigma^2_e}}
})
  \}\\
&\times  [\Phi\{\ \frac{2- (\sum_{k=T+1}^n
q_k^{\prime}
L
q_k
 )/(n-T)}{\sqrt{2\sigmae/(n-T)}}\}\  -
  \Phi\{\frac{-(\sum_{k=T+1}^n
q_k^{\prime}
L
q_k
 )/(n-T)}{\sqrt{2\sigmae/(n-T)}}\}].
 \ee
\end{theorem}

\begin{remark}
To obtain some intuition regarding the marginal likelihood, consider  \\ $-\log \Pi(L; \sigma_e^2,Q)$ as a loss function over $Q$, while ignoring the normalizing constant $\zeta$,
 \be
  - \sum_{k=1}^T
   (q_{k}^{\prime} L
    q_{k})^2 - \frac{1}{n-T}
\big( \sum_{k=T+1}^n
q_k^{\prime}
L
q_k
\big)^2=  - \sum_{k=1}^T
   (q_{k}^{\prime} L
    q_{k})^2 - \frac{m}{n-T}
 \sum_{k=T+1}^n\big(
q_k^{\prime}
L
q_k
\big)^2,
\ee
where $m\in[1,2]$ due to $ \sum x^2_k\le (\sum x_k)^2 \le  2\sum x^2_k$, with $x_k=q_k^{\prime} L q_k \ge 0$ with $L$   being positive semi-definite.
Therefore, the first $T$ factors $(q_k^{\prime} L q_k)^{2}$ have a substantially higher contribution compared to the remaining ones, which is consistent with our modeling focus on the first $T$ eigenvectors.
\end{remark}

Lastly, we show that the proposed non-parametric model of the matrix containing the eigenvectors is posterior consistent. There has been theoretic work on community detection and eigenvector estimation for single graphs, assuming that the
number of vertices $n$ goes to infinity. A fundamental difference here is that we have fixed and bounded $n$ in each graph, but the number of graphs $S$ grows. Hence, a new theoretical approach is required.

In order to avoid a potential discrepancy between the number of spikes in the true model and prescribed model, we use the {\em full} eigen-decomposition for the raw observed $L^{(s)} = W^{(s)} \Omega^{(s)}W^{(s)\prime}$, where $W^{(s)}$ is an orthonormal matrix and $\Omega^{(s)}$ diagonal. Note that $W^{(s)}$ belong to a a Stiefel sub-manifold $\mathcal V^* \subseteq \mathcal V^{n,n}$, with the first column elements being all positive. 
Similarly, for the spiked graph Laplacian we have $\mu_L= Q^{\dagger}\Lambda^{\dagger}Q^{\dagger \prime}$, where $Q^{\dagger}\in \mathcal{V}^*$ and the first $T$ columns equal to parameter $Q$, $\Lambda^{\dagger} = \diag{\lambda_1,\ldots,\lambda_T, \theta,\ldots,\theta}$.

Using $f$ to denote the likelihood, each observed $L^{(s)}=W^{(s)} \Omega^{(s)}W^{(s)\prime}$ can be generated from
\be
 f(W^{(s)}, \Omega^{(s)} \mid Q^\dagger, \Lambda^\dagger)
\propto &        \underbrace{ \etr  \big\{
       \frac{1}{2\sigma_e^2}  Q^\dagger  \Lambda^\dagger Q ^{\dagger\prime}  W^{(s)} \Omega^{(s)}W^{(s)\prime}
           \big\} }_{f(W^{(s)} \mid \Omega^{(s)},Q^\dagger, \Lambda^\dagger)}
           \underbrace{
            \etr \big\{
        -   \frac{1}{4\sigma_e^2}   \big[ \Omega^{(s) }   \Omega^{(s) } + \Lambda^\dagger\Lambda^\dagger \big]
             \big\}}_{ f(\Omega^{(s)} \mid  \Lambda^\dagger)}.
\ee
 The former corresponds to $W^{(s)} \sim\text{Matrix-Bingham}\big[ \Omega ^{(s)},(2\sigma_e^2)^{-1}  Q^\dagger  \Lambda^\dagger Q ^{\dagger\prime} \big]$, in which $Q^\dagger$ serves as the location parameter. 
 
 Therefore, based on a non-parametric mixture prior for $Q^\dagger$, our task is equivalent to showing the consistency of estimating $Q^\dagger\in\mathcal V^*$ under the Matrix-Bingham likelihood. Using the $Q^\dagger$-marginal density 
$
 f_{Q^\dagger}(W^{(s)}) = 
 \int \int f(W^{(s)}, \Omega^{(s)} \mid Q^\dagger, \Lambda^\dagger)    P(\textup d\Lambda^\dagger,{\textup d} \Omega^{(s)}),
$
where $P(.)$ denotes the appropriate measure,
 consider a neighborhood of the true density 
$f_{Q^\dagger,0}$ on the manifold $\mathcal V^{*}$ as
\be
B_\epsilon(f_{Q^\dagger,0}) = \bigg\{f_{Q^\dagger}:\bigg |\int g f_{Q^\dagger} \mu({\textup d} W) - g f_{Q^\dagger,0} \mu( {\textup d} W) \bigg| \le \epsilon, \quad \forall  g\in C_b(\mathcal
V^{*}) \bigg\},
\ee
with $C_b$ denoting the class of continuous and bounded functions, and $\mu(.)$ the Haar measure on $\mathcal V^{*}$. 
Next, we establish that the probability for the posterior
density falling into $B_\epsilon(f_{Q^\dagger,0})$ goes to $1$ as $S\to \infty$.
 \begin{theorem}
[Posterior consistency for the estimated eigenmatrix]
Let $W^{(1)}\ldots W^{(S)}$ be matrices of eigenvectors, whose elements are independently and identically distributed from a distribution with density $f_{Q^\dagger,0}$. Then, for all $\epsilon>0$, as $S\to \infty$,
\be
\Pi \big\{B_\epsilon(f_{Q^\dagger,0})  \mid W^{(1)},\ldots ,W^{(S)}\big\} = \frac{
\int_{B_\epsilon(f_{Q^\dagger,0}) } \prod_{s=1}^S f_{Q^\dagger}(W^{(s)}) \Pi(\textup{d}f)
}{
\int \prod_{s=1}^S f_{Q^\dagger}(W^{(s)}) \Pi(\textup{d}f)
}
\to 1
\; a.s.  Pf_{Q^\dagger,0}^{\infty},
\ee
with $Pf_{Q^\dagger,0}^{\infty}$ the true probability measure for $(W^{(1)},W^{(2)},\ldots)$.
\end{theorem}

\section{Performance Evaluation based on Synthetic Data}
\subsection{Impact of Different Noise Levels on a Single Graph}
We first examine the effects of noise on the estimation of the communities in a single graph. We generate a weighted graph comprising of $60$ vertices and three communities of size 10, 20 and 30 vertices, respectively. To avoid directly using the proposed model to generate data, we simulate each edge within the communities as a Bernoulli event with probability $0.5$, and then add to it Gaussian noise $\text{No}(0,\xi^2)$ to the adjacency matrix with varying $\xi^2$. Hence, the spiked graph Laplacian model serves as a working model for the true network generating mechanism.

\begin{figure}[H]
 \begin{subfigure}[t]{0.32\textwidth}
 \centering
  \captionsetup{justification=centering}
  \caption*{\footnotesize Low-noise graph\\ (observed spectral gap 0.6)}
       \includegraphics[width=1\linewidth]{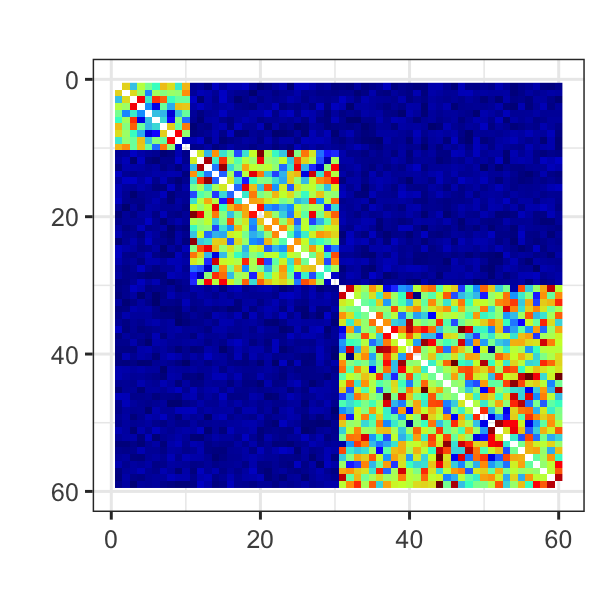}
   \end{subfigure}
 \begin{subfigure}[t]{0.32\textwidth}
 \centering
 \captionsetup{justification=centering}
  \caption*{\footnotesize Medium-noise graph\\  (observed spectral gap 0.3)}
       \includegraphics[width=1\linewidth]{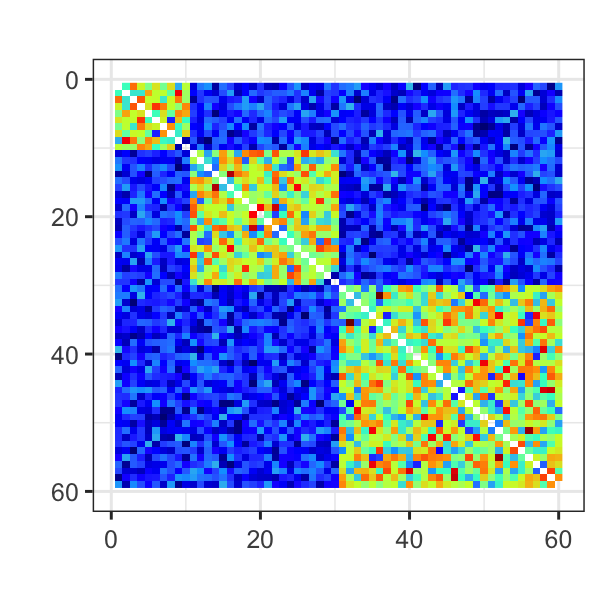}
 \end{subfigure}
  \begin{subfigure}[t]{0.32\textwidth}
 \centering
  \captionsetup{justification=centering}
  \caption*{\footnotesize High-noise graph\\ (observed spectral gap 0.05)}
       \includegraphics[width=1\linewidth]{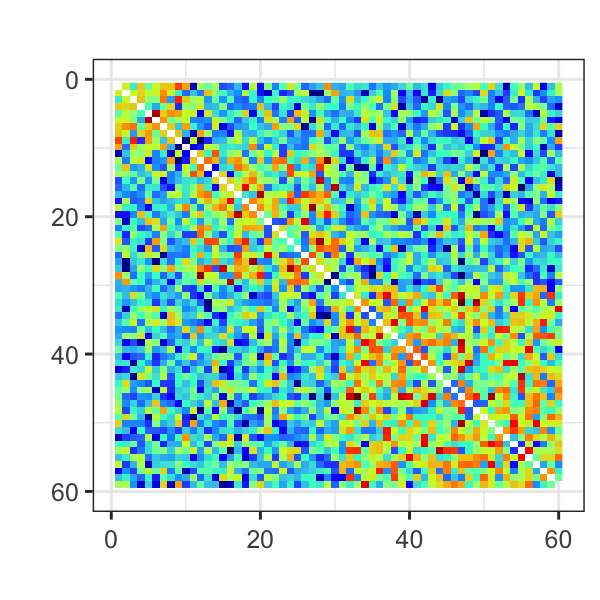}
   \end{subfigure}
 \begin{subfigure}[t]{0.32\textwidth}
 \centering
       \includegraphics[width=2.9in, height=2in]{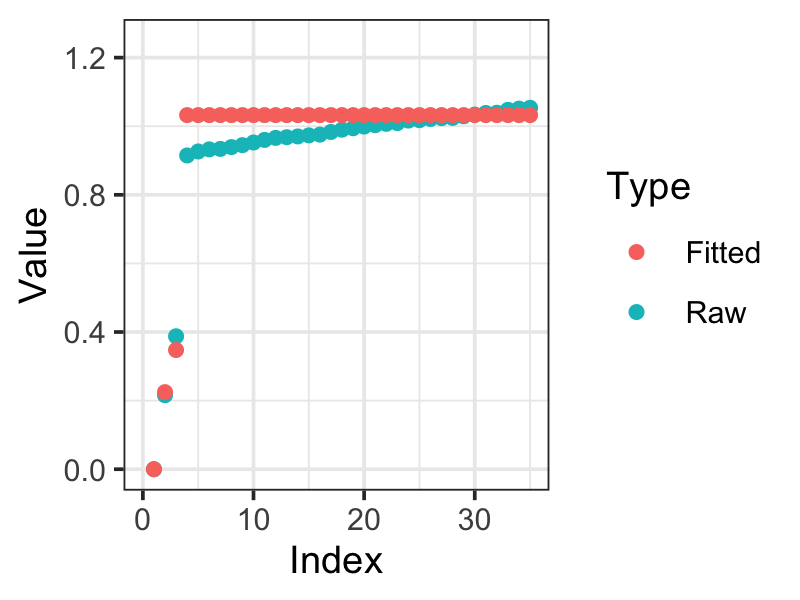}
   \end{subfigure}
 \begin{subfigure}[t]{0.32\textwidth}
 \centering
       \includegraphics[width=2.9in, height=2in]{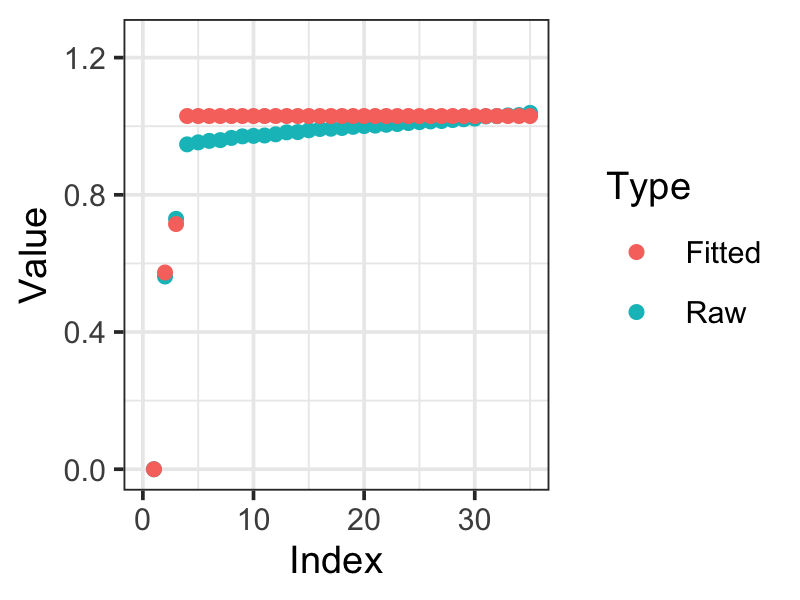}
 \end{subfigure}
  \begin{subfigure}[t]{0.32\textwidth}
 \centering
       \includegraphics[width=2.9in, height=2in]{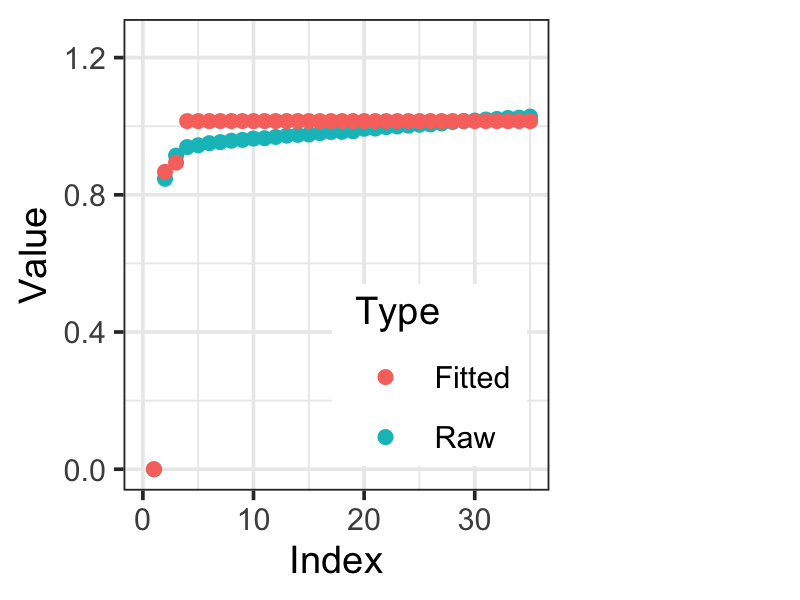}
   \end{subfigure}
   \caption{ Three simulated graphs with different degree of noise, corresponding to different spectral gaps in the eigenvalues (for clarity, we show the first $35$ eigenvalues out of $60$). Comparing the eigenvalues produced by the direct decomposition of the raw Laplacian (cyan), and the ones by spiked Laplacian model (red), the latter has a clearly larger spectral gap between the third and fourth, corresponding to better separation between signal and noise. \label{fig:spectral_gap}}
 \end{figure}

As shown in Figure~\ref{fig:spectral_gap}, the $3$-community structure can be visualized by the spectral gap between the third and fourth eigenvalues. As the noise increases, the gap diminishes, making it more difficult to separate the communities.

The spiked Laplacian model has a ``lifting effect" on the fourth eigenvalue (shown in red in Figure~\ref{fig:spectral_gap}). This is due to the flat structure imposed, effectively replacing the fourth eigenvalue $\hat\lambda_4$ by $\theta\approx(\sum_{k=4}^{60}\hat\lambda_k)/57$ with $\hat\lambda_k>\hat\lambda_4$ for $k>4$. Consequently, it leads to an increase in the spectral gap, compared to a direct eigendecomposition of the graph Laplacian (shown in cyan). Practically, this leads to improved accuracy in finding the community labels, as shown in Table~\ref{tab:spectral_gap_nmi}. 
This phenomenon can be viewed as a result of rank regularization on the Laplacian matrix; \cite{le2018concentration} discussed similar effects under a slightly different regularization in spectral clustering.

\begin{table}[H]
      \footnotesize
  \centering
      \begin{tabular}{ l | c c c c c }
              \hline
      Observed Spectral Gap ($\hat\lambda_4-\hat\lambda_3$)&   0.6 & 0.3 & 0.1 & 0.05 & 0.01  \\
          \hline
      Spiked Laplacian & ($1 \pm 0$)  & ($0.95 \pm 0.05$)  & ($0.88 \pm 0.09$) & ($0.58 \pm 0.20$) & ($0.40 \pm 0.14$) \\
      Observed Laplacian & ($1 \pm 0$)  &  ($0.91 \pm 0.07$) &  ($0.78 \pm 0.05$) & ($0.42 \pm 0.20$) & ($0.36 \pm 0.25$) \\
      \hline
      \end{tabular}
      \caption{The spiked Laplacian model has higher accuracy in recovering community labels, comparing to direct decomposition of the observed Laplacian. The results are calculated by clustering the second and third eigenvectors into three groups, then comparing with the ground truth labels to compute normalized mutual information (NMI). Mean $\pm$ standard deviation is reported based on 50 times of experiments. The higher the NMI, the higher the accuracy.
      \label{tab:spectral_gap_nmi}}
  \end{table}

\subsection{Impact of Graph Size on Community Detection}

\begin{table}[H]
\centering
    \begin{tabular}{ l | c c c c}
            \hline
            $n$ &   100 & 300 & 500 & 1000  \\
        \hline
            Spiked Laplacian Model & ($0.84 \pm 0.15$)  & ($0.90 \pm 0.04$)  & ($0.86 \pm 0.04$) & ($1\pm 0$)  \\
     Stochastic Block Model & ($0.65 \pm 0.23$) & ($0.84 \pm  0.09$) & ($0.87 \pm 0.04$) & ($1 \pm 0$)\\
         Bayesian SBM & ($0.70 \pm 0.14$)  & ($0.83 \pm 0.08$)  & ($0.88 \pm 0.05$)  & ($1 \pm 0$)\\
    \hline
    \end{tabular}
    \caption{At small $n$, the spiked Laplacian model has higher accuracy in estimating the community labels. Mean $\pm$ standard deviation is reported based on 50 times of experiments. The higher the NMI, the higher the accuracy.  \label{tab:varying_n}}
\end{table}
Next, we evaluate the effects of a different number of vertices ($n$) on community detection. We adopt a similar $3$-community setting as in the previous subsection, retaining the community size ratio as  $1:2:3$, and increase the total number of vertices. We calculate the normalized mutual information that compares the estimated community labels and the ground truth, using estimates produced by the spiked graph Laplacian model, stochastic block model using the spectral clustering algorithm \citep{ng2002spectral} and Bayesian stochastic block model using a Gibbs sampler based on the model in \cite{van2018bayesian}.

\begin{figure}[H]
 \begin{subfigure}[t]{0.3\textwidth}
 \centering
       \includegraphics[width=1\linewidth, height=1.3in]{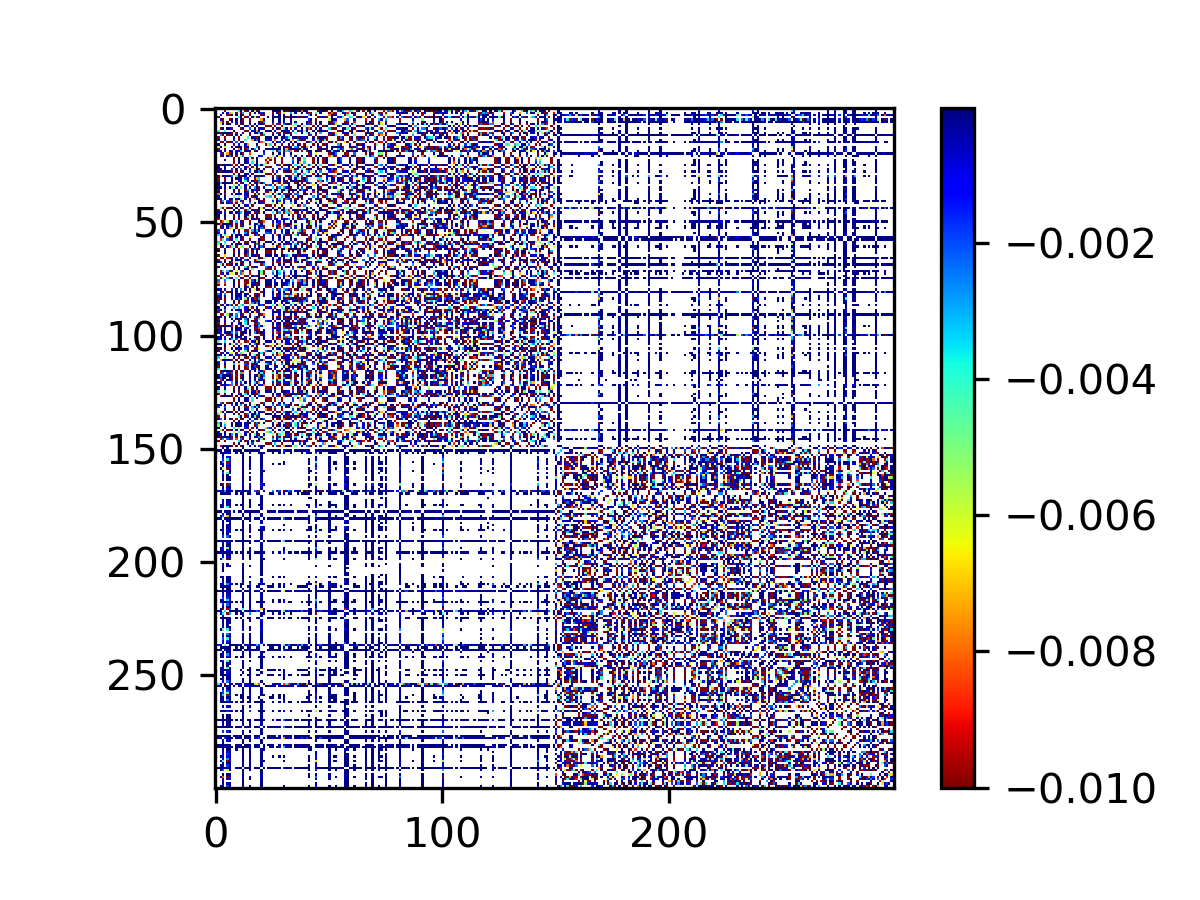}
       \caption{\footnotesize Observed Laplacian $L$ based on the simulated graph.}
 \end{subfigure}
 \begin{subfigure}[t]{0.3\textwidth}
 \centering
       \includegraphics[width=1\linewidth, height=1.3in]{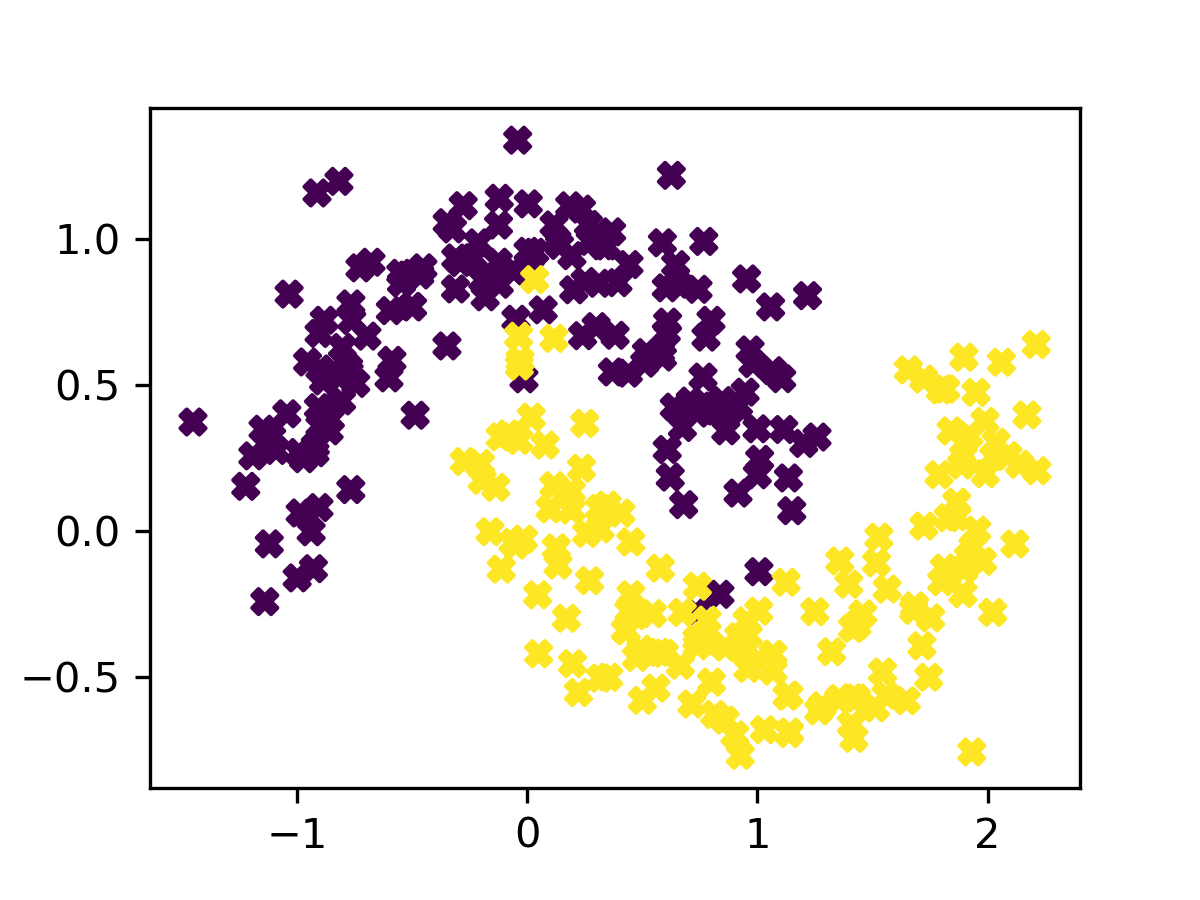}
       \caption{\footnotesize The latent positions $y_i$ used to generate the graph, colored by the true labels.}
   \end{subfigure}
  \begin{subfigure}[t]{0.3\textwidth}
 \centering
       \includegraphics[width=1\linewidth, height=1.3in]{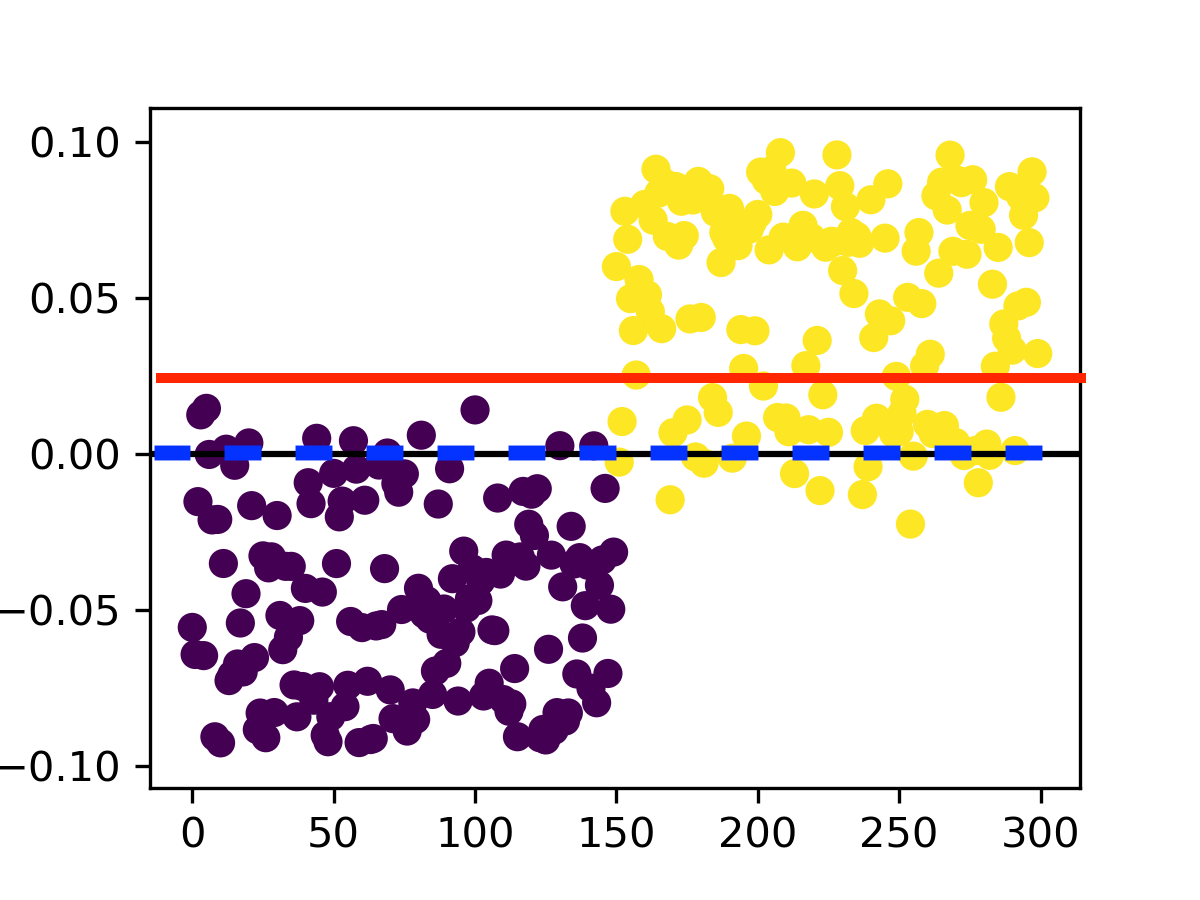}
       \caption{\footnotesize $q_2(i)$ vs vertex index. Two communities are classified by the K-means (red) or sign-based algorithm (blue).}
   \end{subfigure}
      \begin{subfigure}[t]{0.47\textwidth}
 \centering
        \includegraphics[width=1\linewidth, height=1.5in]{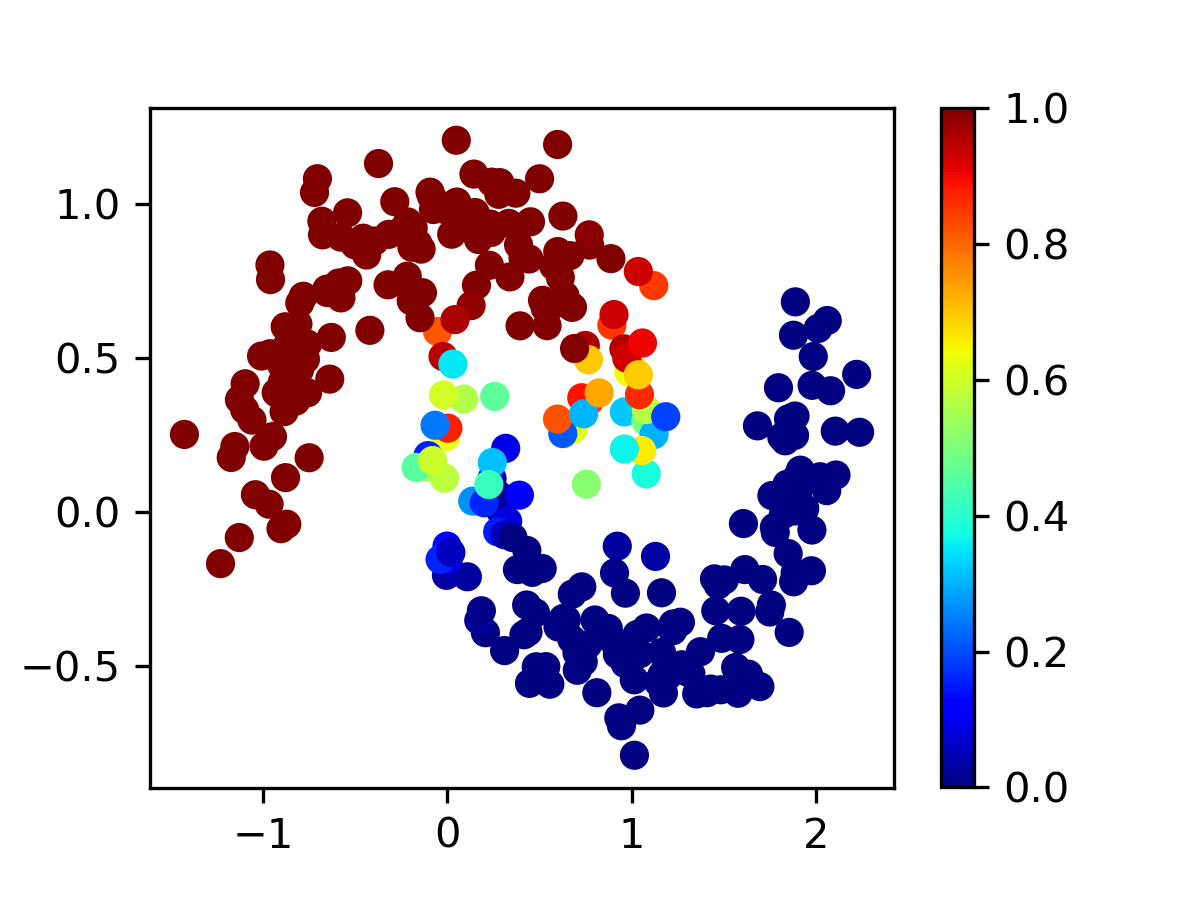}
       \caption{\footnotesize Spiked Laplacian model correctly estimated the uncertainty $\text{pr}(c_i=1)$, based on  sign-partitioning of {\em each} posterior sample $q_2$.}
   \end{subfigure}
   \qquad
     \begin{subfigure}[t]{0.47\textwidth}
  \centering
       \includegraphics[width=1\linewidth, height=1.5in]{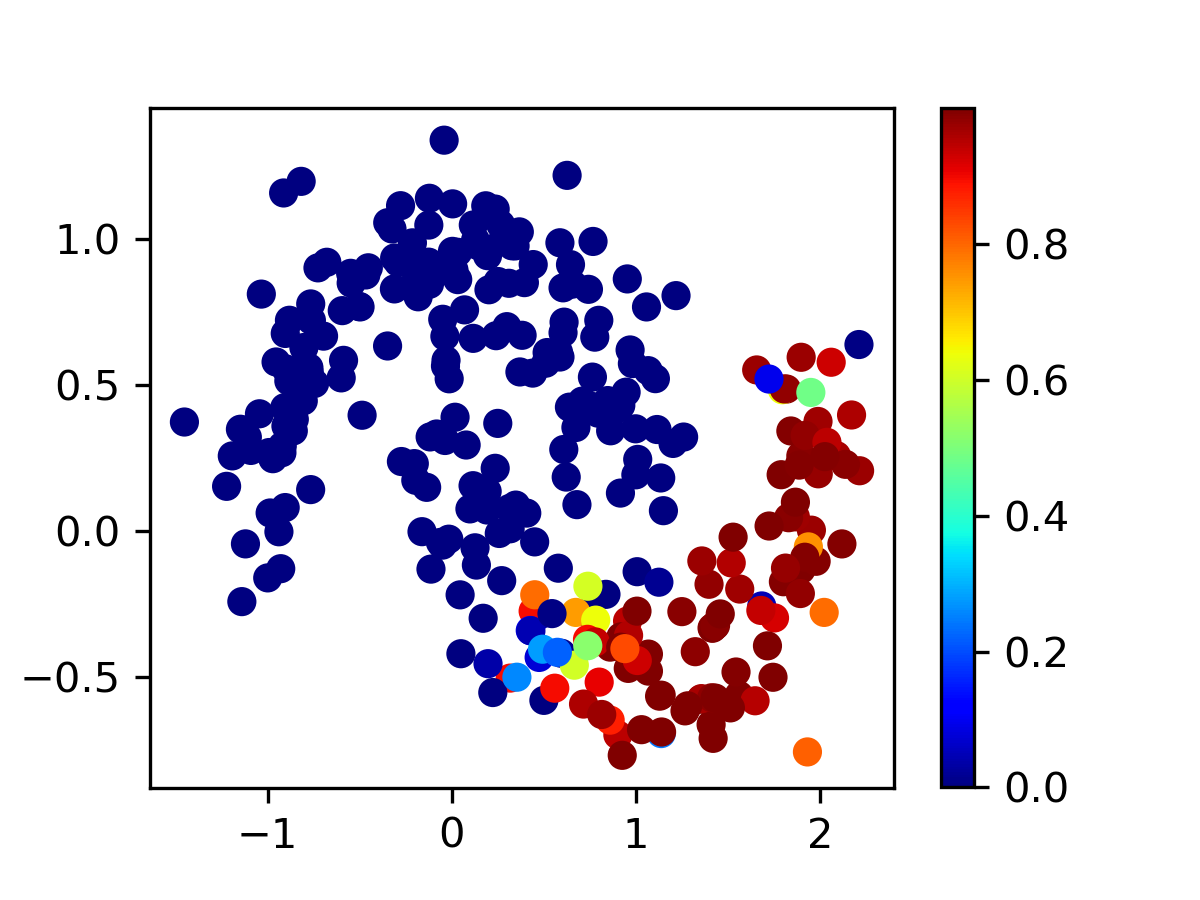}
       \caption{\footnotesize Applying Gaussian mixture model on {\em only one} sample of $q_2(i)$ (as in the stochastic block model) under-estimates the uncertainty  $\text{pr}(c_i=1)$.}
   \end{subfigure}
\quad
 \caption{
 Illustration of uncertainty quantification by the spiked graph Laplacian model.
  \label{fig:sim_uq}
  }
 \end{figure}
As shown in Table~\ref{tab:varying_n}, for large $n\ge 500$, there are almost no differences in terms of the point estimate accuracy. However, at smaller vertex sizes $n$, the spiked graph Laplacian model exhibits clearly superior performance.

Next, we empirically show that the advantage for small $n$ is attributable to more accurate uncertainty quantification. For a more intuitive illustration of this issue, we generate a $2$-community graph using a latent position model --- we first sample latent $y_i$ near two manifolds [Figure~\ref{fig:sim_uq}, Panel (b)], then compute the pairwise similarity between latent positions [$A_{i,j}=\exp(-10\|y_i-y_j\|_2)$], and use it as the edge weight. Clearly, most of the uncertainty is located in the center of the adjacency matrix, where the manifolds get close.

Panel (c) plots one sample of $\vec q_2$. The sign-based partition used by the spiked graph Laplacian model has a default decision boundary at the zero line (in blue). Applying this bi-partitioning on each posterior sample of $\vec q_2$, it leads to an accurate uncertainty quantification [Panel (d)]. Comparatively, in the estimation of stochastic block model, one applies K-means or a Gaussian mixture model on {\em one} sample of $\vec q_2$ (based on the direct eigendecomposition of $L$), which could result in a severe underestimation of the uncertainty, as shown in Panel (e).
\subsection{Accommodating Heterogeneity in a Collection of Graphs}

In this experiment, we deal with multiple graphs comprising of 300 vertices each, whose adjacency matrices exhibit heterogeneity.
We first generate a set of five possible community structures, each represented by a binary matrix (denoted by $W^{(l)}$) of size $300\times 6$; each row has one $1$ and five $0$'s, encoding the ground truth  of the community labels in  $1,\ldots,6$. To generate a graph, we randomly draw  one of five patterns as $\tilde W^{(s)}$ and a non-negative random vector $\tilde \Lambda$, producing its adjacency matrix by $A^{(s)}=\tilde W^{(s)} \tilde \Lambda \tilde W^{(s) \prime} +  \mathcal{\tilde E}^{(s)}$, with $\mathcal{\tilde E}^{(s)}$ being a Gaussian noise matrix and  $\tilde e^{(s)}_{i,j} =\tilde e^{(s)}_{j,i}  \sim\No(0,1)$.

We compare the performance of the proposed model against several popular alternatives: (1) simple averaging of all graphs followed by the use of a stochastic block model,
 (2) co-regularized stochastic block model/spectral clustering \citep{kumar2011co}, (3) clustering the graphs into five groups, and applying the stochastic block model in each group, (4) independent stochastic block model for each graph. The first two competitors produce only one partitioning, while the latter two accommodate the heterogeneity.

 We compute two benchmark scores: the normalized mutual information (NMI), reflecting the similarity between the estimated community labels to the ground truth in each graph; and the Root Mean Squared Error between the individual $L^{(s)}$ and the smoothed $\hat L^{(s)}$, as the goodness of fit criterion.
 
\begin{table}[H]
\centering
    \begin{tabular}{ l | c c }
            \hline
      Benchmark Scores &  NMI (higher is better)& RMSE ($\times 10^{-3}$, lower is better)   \\
        \hline
     Spiked Laplacian Graphs & $0.85\pm 0.04$  & $1.9 \pm 0.2$  \\
 Average+SBM & $0.21 \pm 0.15$ & $9.2 \pm 2.5$  \\
Co-regularized SBM & $0.25 \pm 0.11$  & $10.2 \pm 4.5$    \\
Clustering Graphs + SBMs & $0.67\pm 0.24$  &  $5.5 \pm 1.5$ \\
Individual SBMs & $0.45\pm 0.13$ & $1.2 \pm 0.2$ \\
    \hline
    \end{tabular}
    \caption{Benchmark of the fitting models to a population of heterogeneous graphs \label{tab:multiple_graphs}. When computing the RMSE, for the Spiked Laplacian Graphs, we obtain $\hat L^{(s)}$ from the spiked representation taking individual $\kappa^{(s)}$ as the truncated dimension, averaging over the posterior sample; for the other four,
    we define $\hat L^{(s)}$ as the truncated spectral representation $\hat Q\hat \Lambda \hat Q$ with $(\hat Q, \hat\Lambda)$ corresponding to the top $6$ dimensions (as the ground truth dimension for data generation).}
\end{table}

As shown in Table~\ref{tab:multiple_graphs}, our proposed model has the highest accuracy in estimating the community labels, followed by the two-stage estimator that clusters the graphs first then partitions the vertices via the stochastic block model. The performance of individual stochastic block models is much worse, likely due to it does not borrow information among graphs, and the vertex number is not too large. In the goodness-of-fit, the individual stochastic block models have the best score as it has the highest flexibility; our model has a slightly larger error; however, it is much lower than the other competitors.

\subsection{Robustness to Over-specified $T$}
Lastly, we examine if the proposed model can handle an over-specified $T$, when it is larger than necessary.  We focus on the following two issues:
(i) whether the posterior sample of $\eta$ can successfully identify redundant $\lambda_k$'s;
 (ii) whether a misspecified $T$ affects the estimation of the first few eigenvalues.

\begin{figure}[H]
 \begin{subfigure}[t]{0.45\textwidth}
 \centering
       \includegraphics[width=6cm, height=4cm]{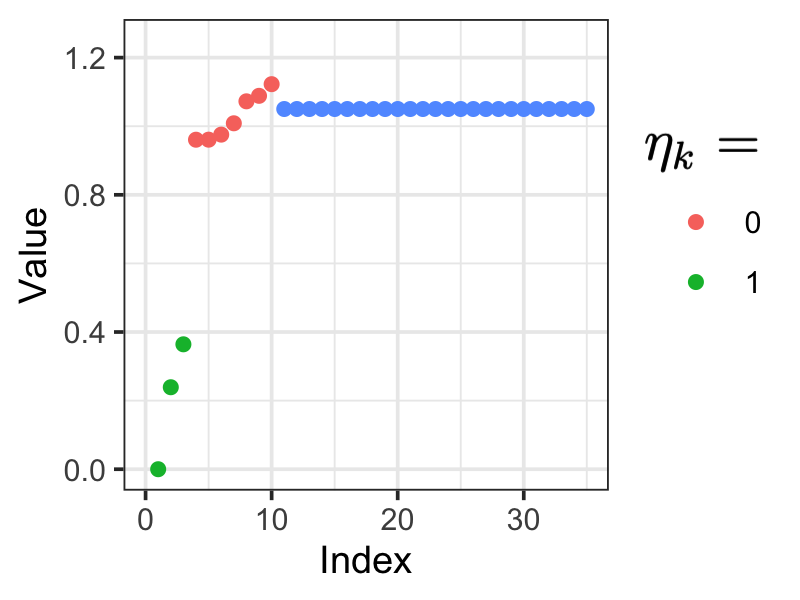}
       \caption{\small Eigenvalues estimated with $T=10$.}
 \end{subfigure}
 \begin{subfigure}[t]{0.45\textwidth}
 \centering
       \includegraphics[width=6cm, height=4cm]{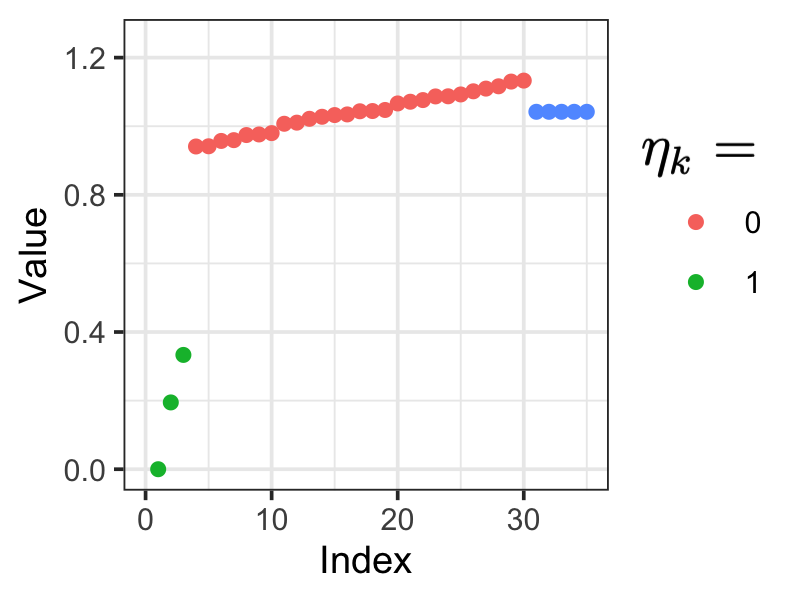}
       \caption{\small Eigenvalues estimated with $T=30$.}
   \end{subfigure}
 \caption{Simulation showing the first few small eigenvalues are almost unaffected by an overly large $T$, and the variable $\eta_k$ successfully identifies the redundant $\lambda_k$.
  \label{fig:overly_large_K}
  }
 \end{figure}
 \vspace{-6mm}

We use the same single graph setup to generate graphs with three communities, except we now set $T=10$ and $T=30$. As shown in Figure~\ref{fig:overly_large_K}, 
the posterior of $\eta_k$ successfully finds all unnecessary $\lambda_k$'s, as indicated by $\eta_k=0$. Further, there is almost no difference in the estimates of the first few eigenvalues $\lambda_1,\lambda_2,\lambda_3$. On the other hand, we should clarify that if the spectral gap is small, $\eta_k$ will more likely be assigned to $0$; this is an expected behavior indicating there is a large loss if we still want to partition the graph.

 \section{Data Application: Characterizing Heterogeneity in a Human Working Memory Study}
 We employ the proposed spiked graph Laplacian model on data obtained from a neuroscience study on working memory, focusing on human brain functional connectivity \citep{hu2019working}. The study involved 1,329 brain scans, wherein each subject in the study was asked to do the Sternberg verbal working memory task, which involved memorizing a list of six numbers, followed by a memory retrieval task that requires the subject to answer if a number was among the six shown earlier. Electroencephalogram (EEG) signals were obtained from $128$ electrode channels placed over each subject's head, and subsequently, a $128\times 128$ connectivity network is estimated during the retrieval task period. Each network has weighted edges taking values in the 
 $[0,1]$ interval. 
 
\begin{figure}[H]
  \begin{subfigure}[t]{0.32\textwidth}
  \centering
        \includegraphics[width=1\linewidth,height=1.8in]{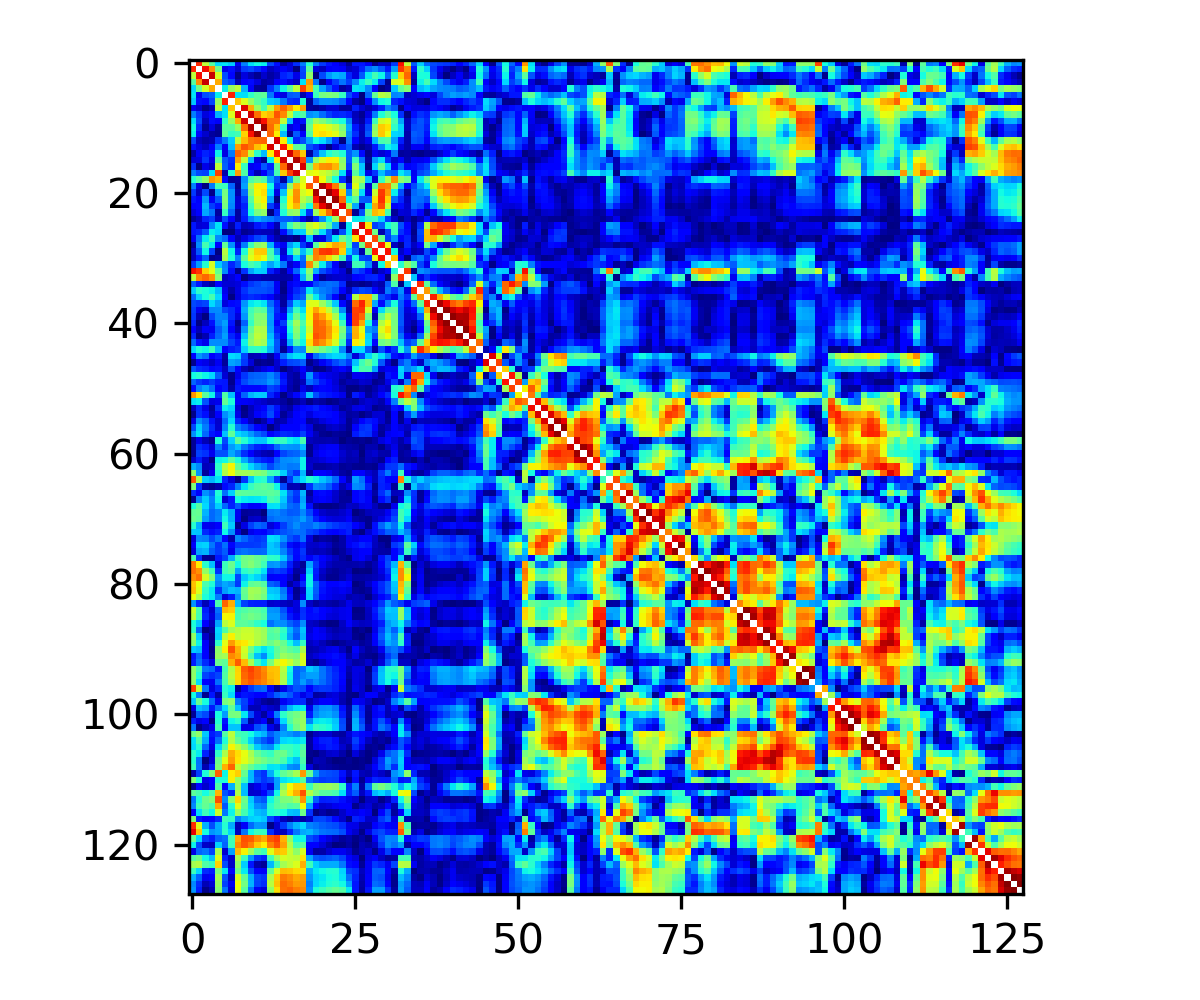}
  \end{subfigure}
 \begin{subfigure}[t]{0.32\textwidth}
  \centering
        \includegraphics[width=1\linewidth,height=1.8in]{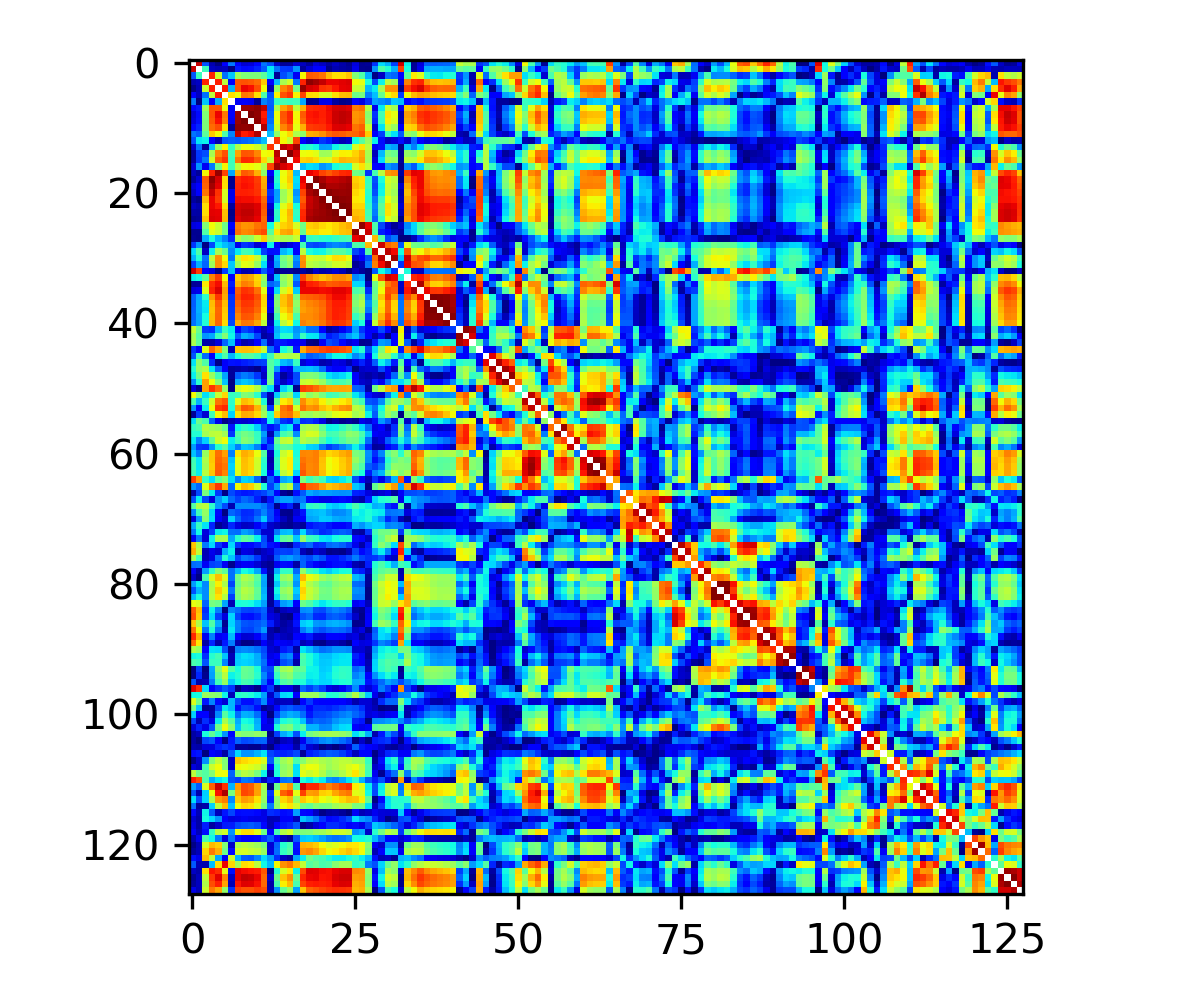}
  \end{subfigure}
  \begin{subfigure}[t]{0.32\textwidth}
  \centering
        \includegraphics[width=1\linewidth,height=1.8in]{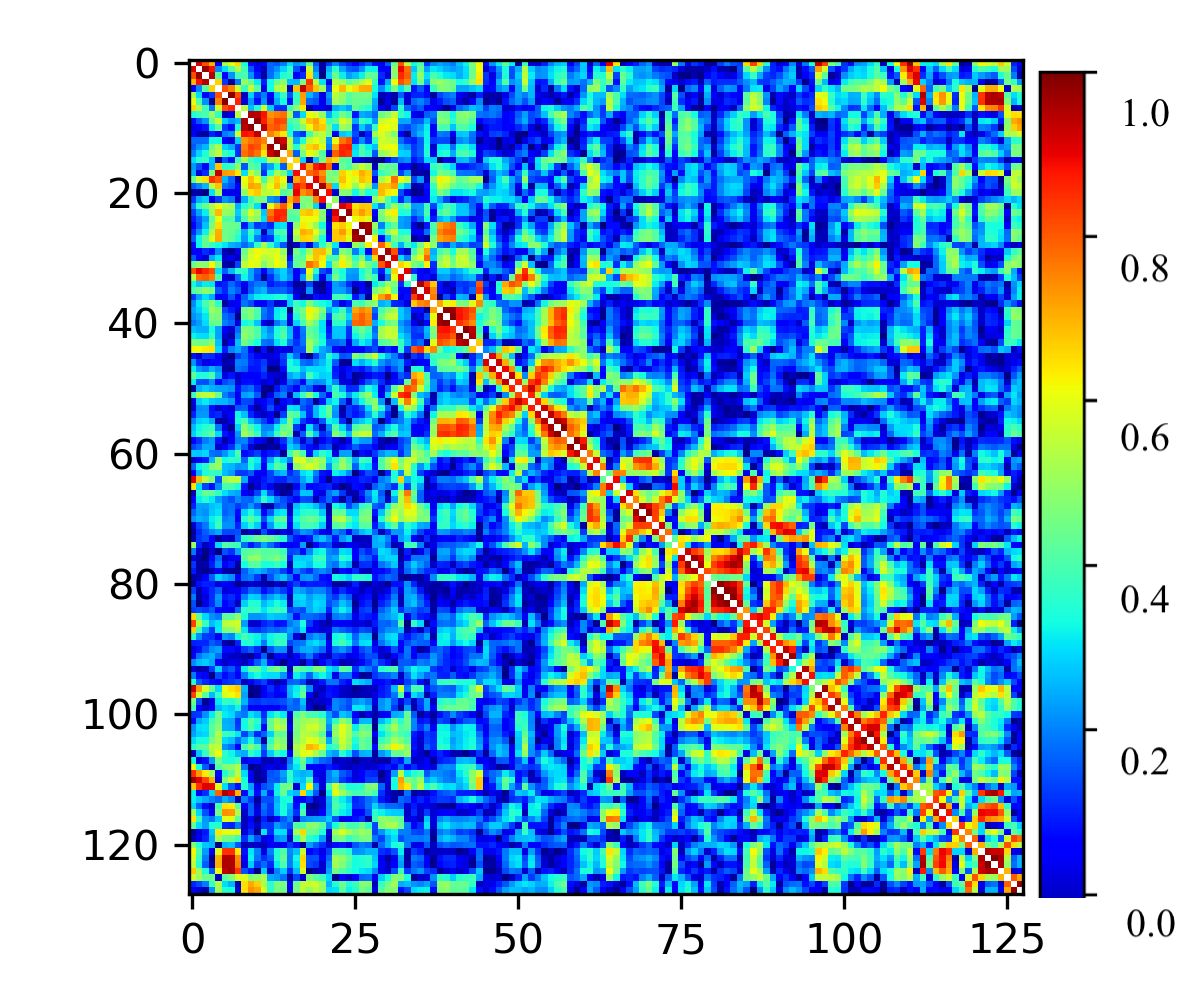}
  \end{subfigure}
   \caption{\small Brain functional connectivity adjacency matrices of three individuals undertaking the memory retrieval task. 
       A significant degree of heterogeneity can be observed.
       \label{fig:working_memory_data}}
         \begin{subfigure}[t]{0.32\textwidth}
  \centering
        \includegraphics[width=1\linewidth,height=1.8in]{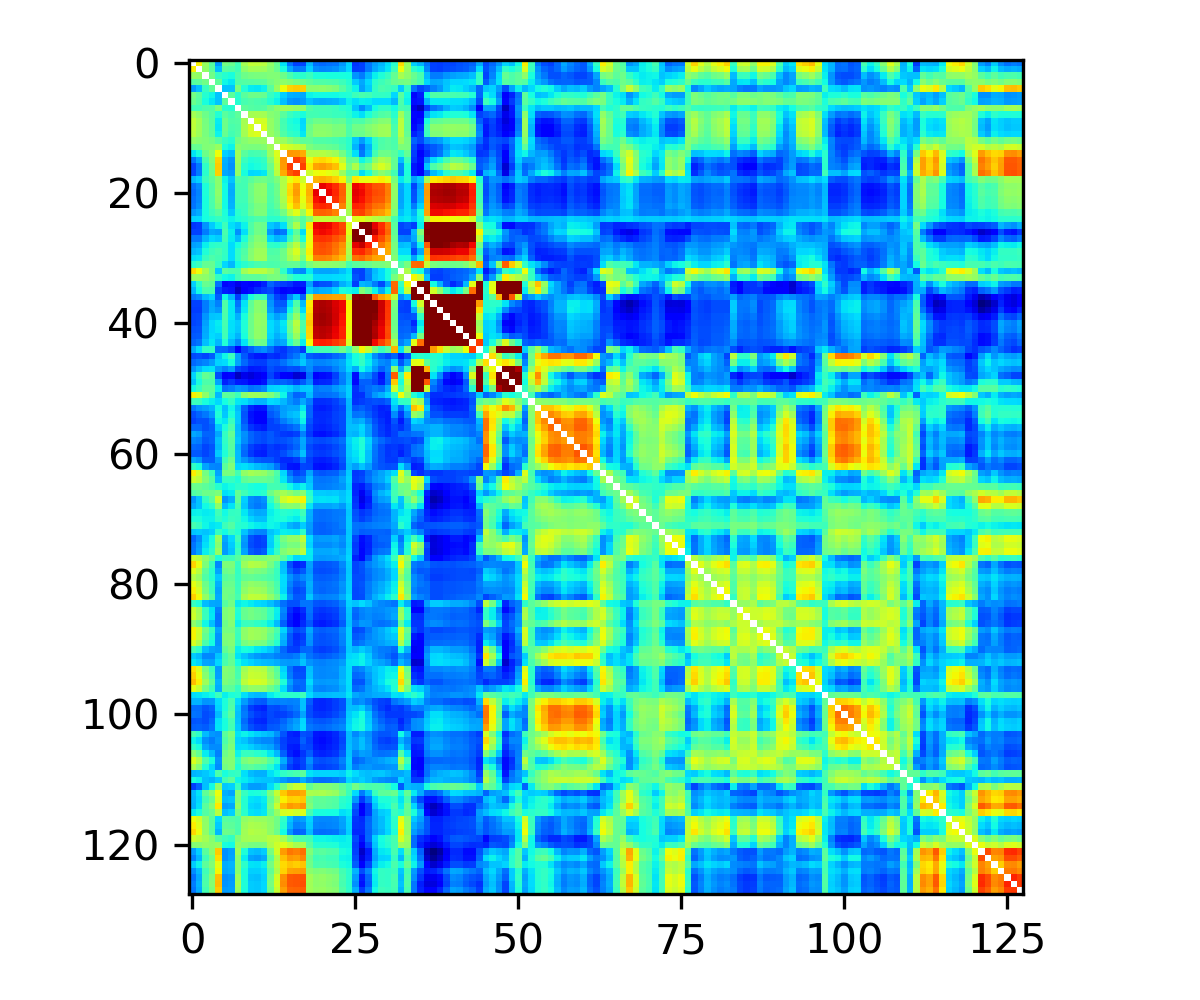}
  \end{subfigure}
  \begin{subfigure}[t]{0.32\textwidth}
  \centering
        \includegraphics[width=1\linewidth,height=1.8in]{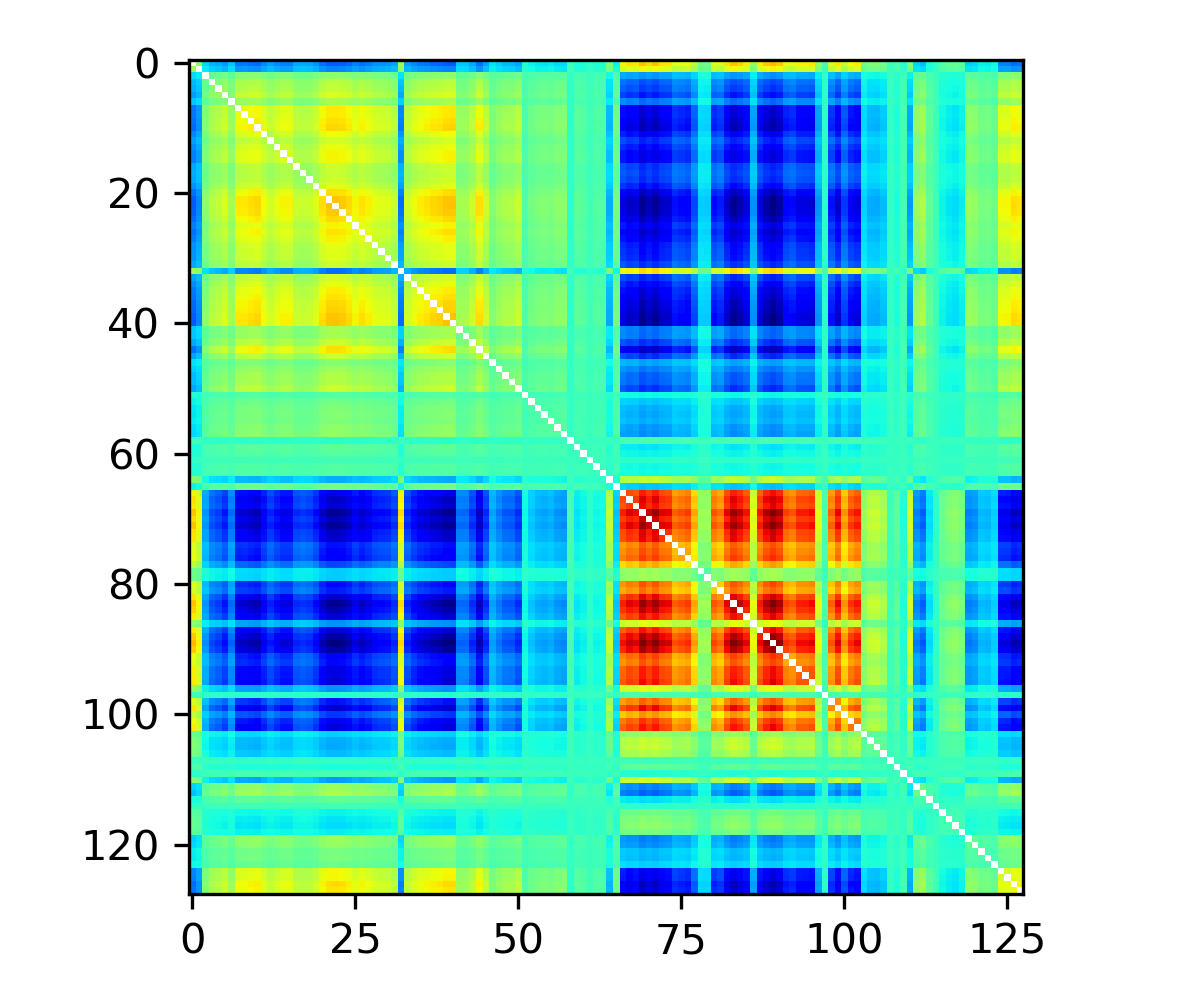}
  \end{subfigure}
 \begin{subfigure}[t]{0.32\textwidth}
  \centering
        \includegraphics[width=1\linewidth,height=1.8in]{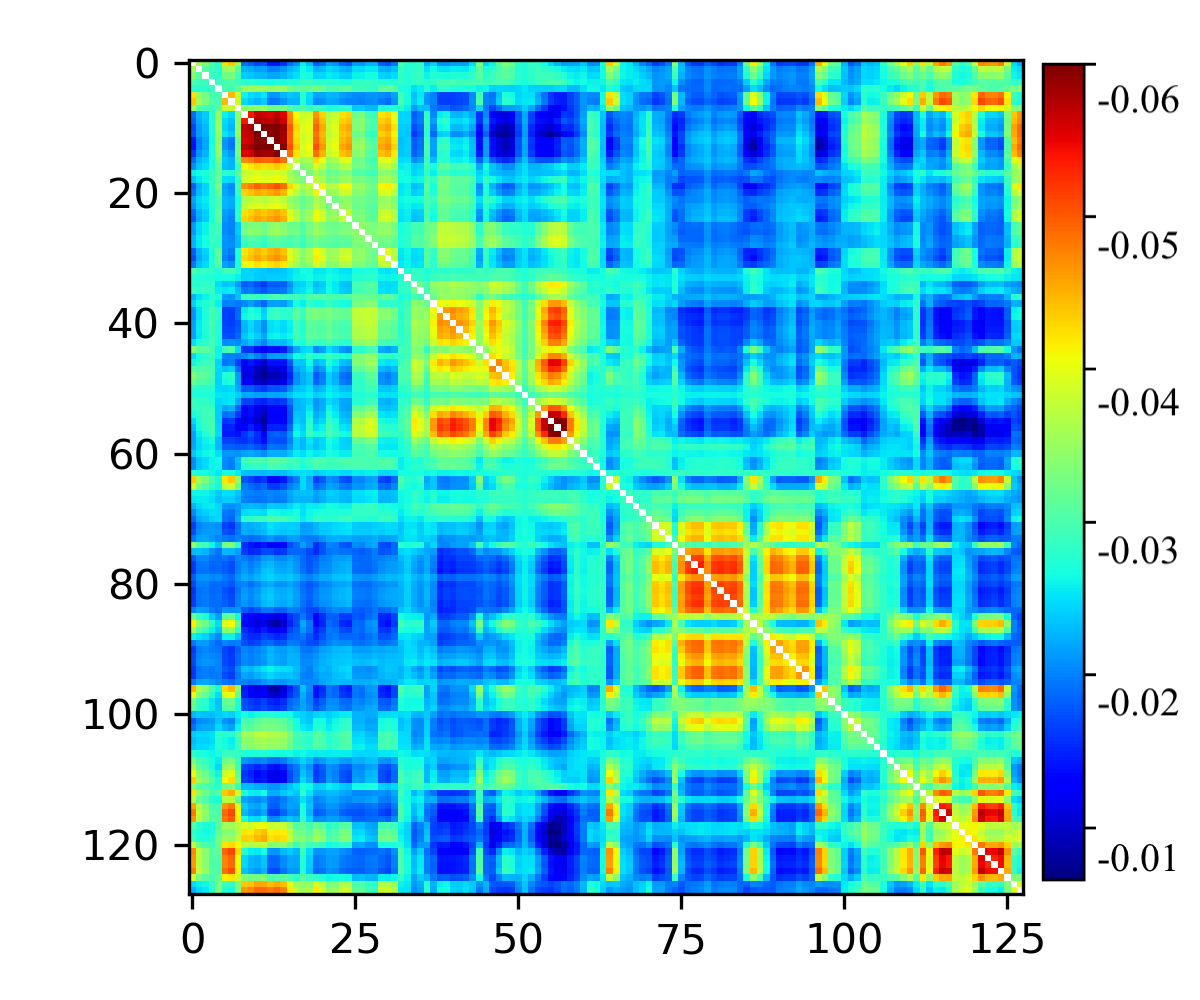}
  \end{subfigure}
   \caption{Fitted Laplacian shows the structure underneath each raw connectivity matrix.\label{fig:fitted-Laplacian}}
  \end{figure}

\begin{figure}[H]
  \begin{subfigure}[t]{0.4\textwidth}
  \centering
        \includegraphics[width=2in, height=2in]{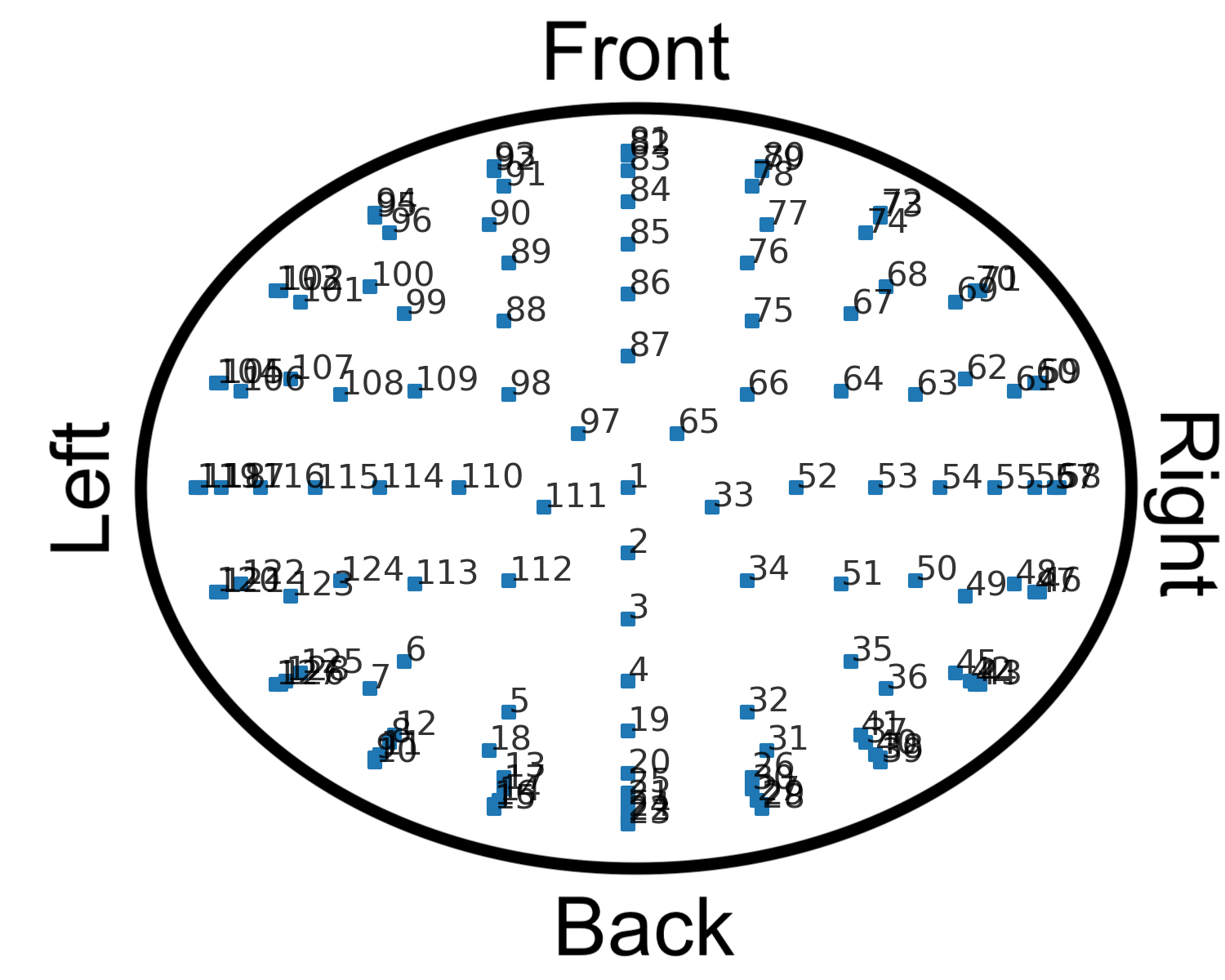}
               \caption{\small Coordinates of the EEG sensors, viewed from the top of the head.}
  \end{subfigure}
   \begin{subfigure}[t]{0.5\textwidth}
        \includegraphics[width=1\linewidth, height=2in]{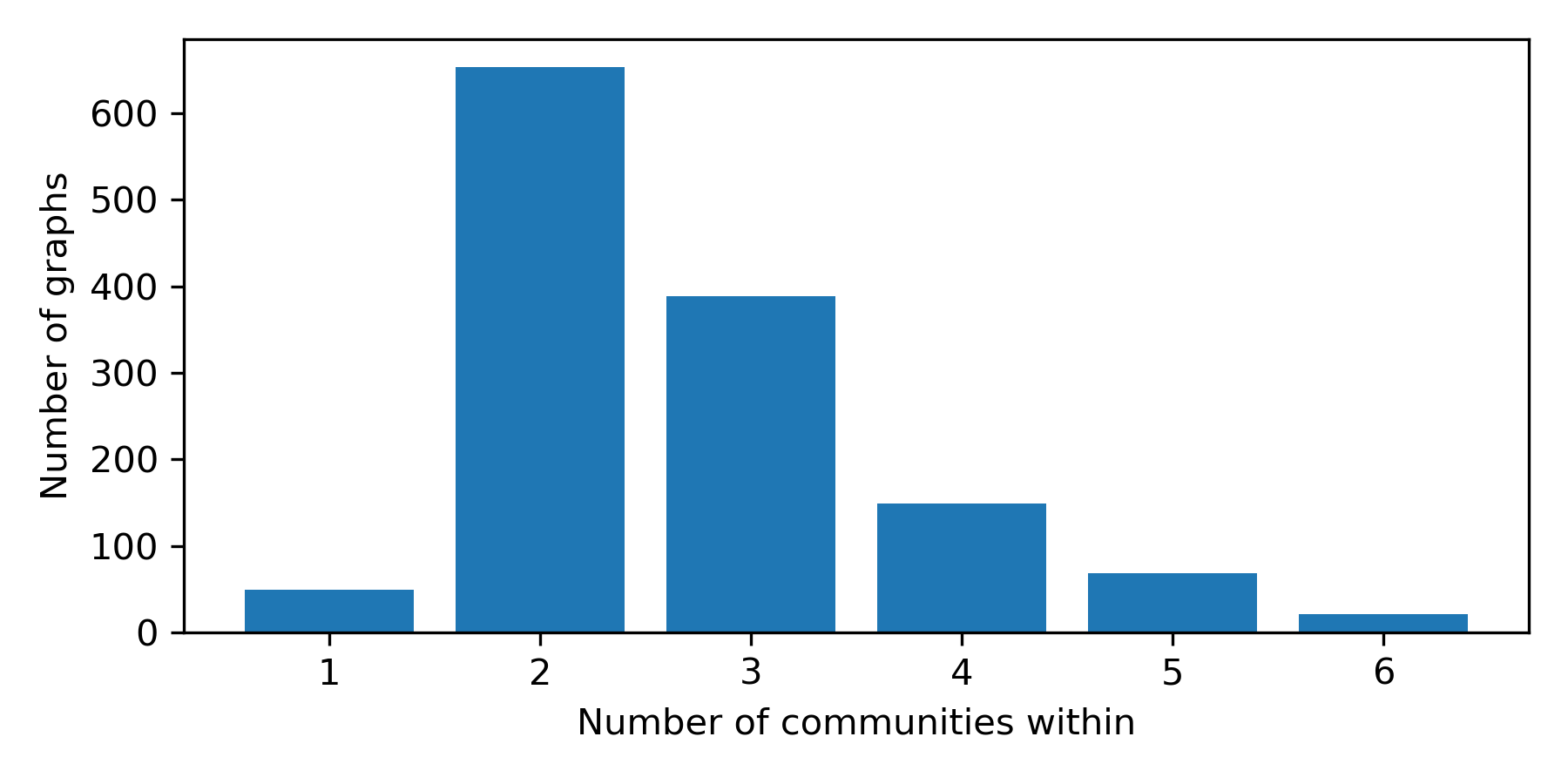}
               \caption{\small Histogram of the number of communities in all subjects.}
  \end{subfigure}
  \begin{subfigure}[t]{0.32\textwidth}
        \includegraphics[width=1\linewidth, height=2in]{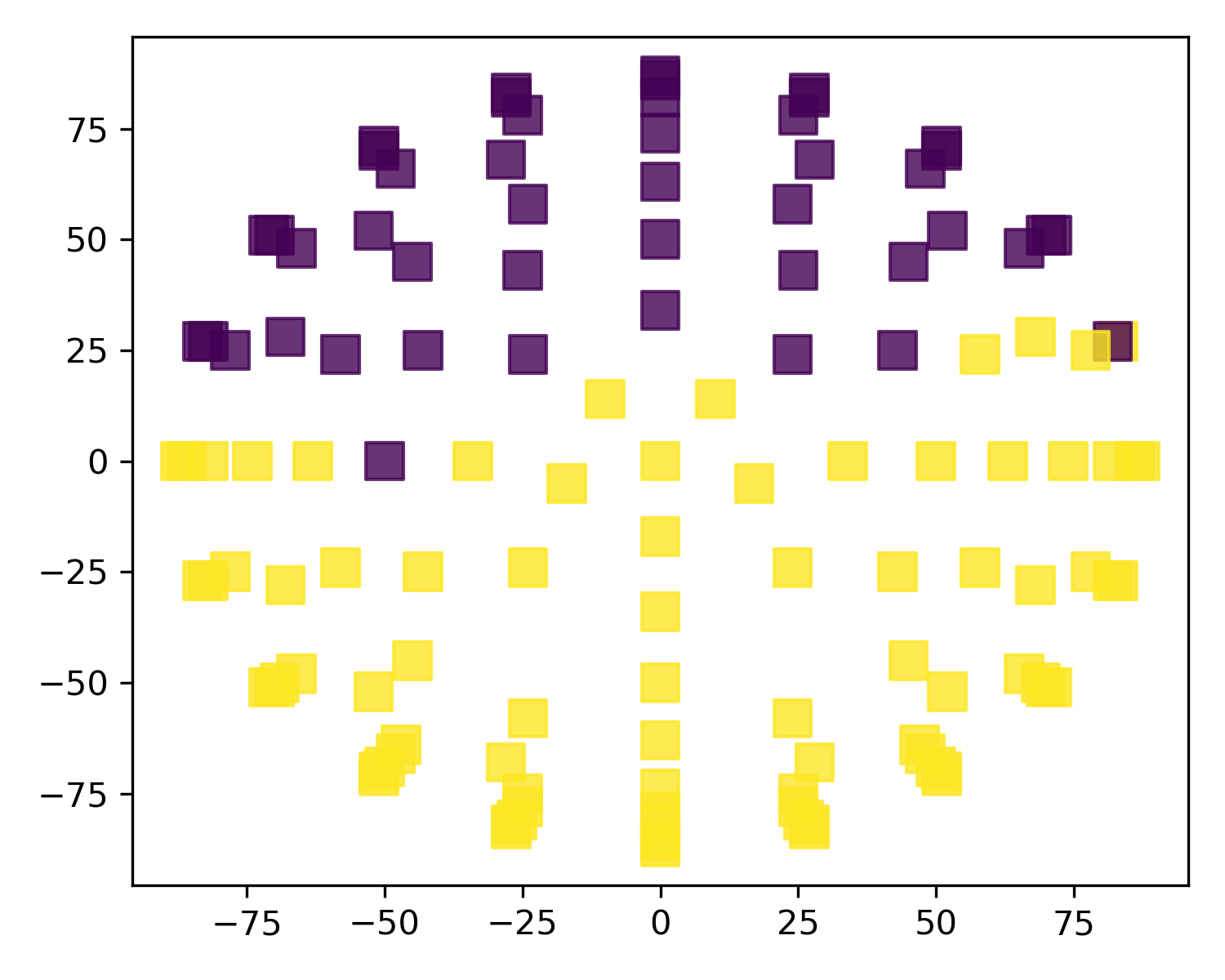}
                      \caption{\small The subject has two communities, with the larger one near the back of the head.}
  \end{subfigure}
  \begin{subfigure}[t]{0.32\textwidth}
        \includegraphics[width=1\linewidth, height=2in]{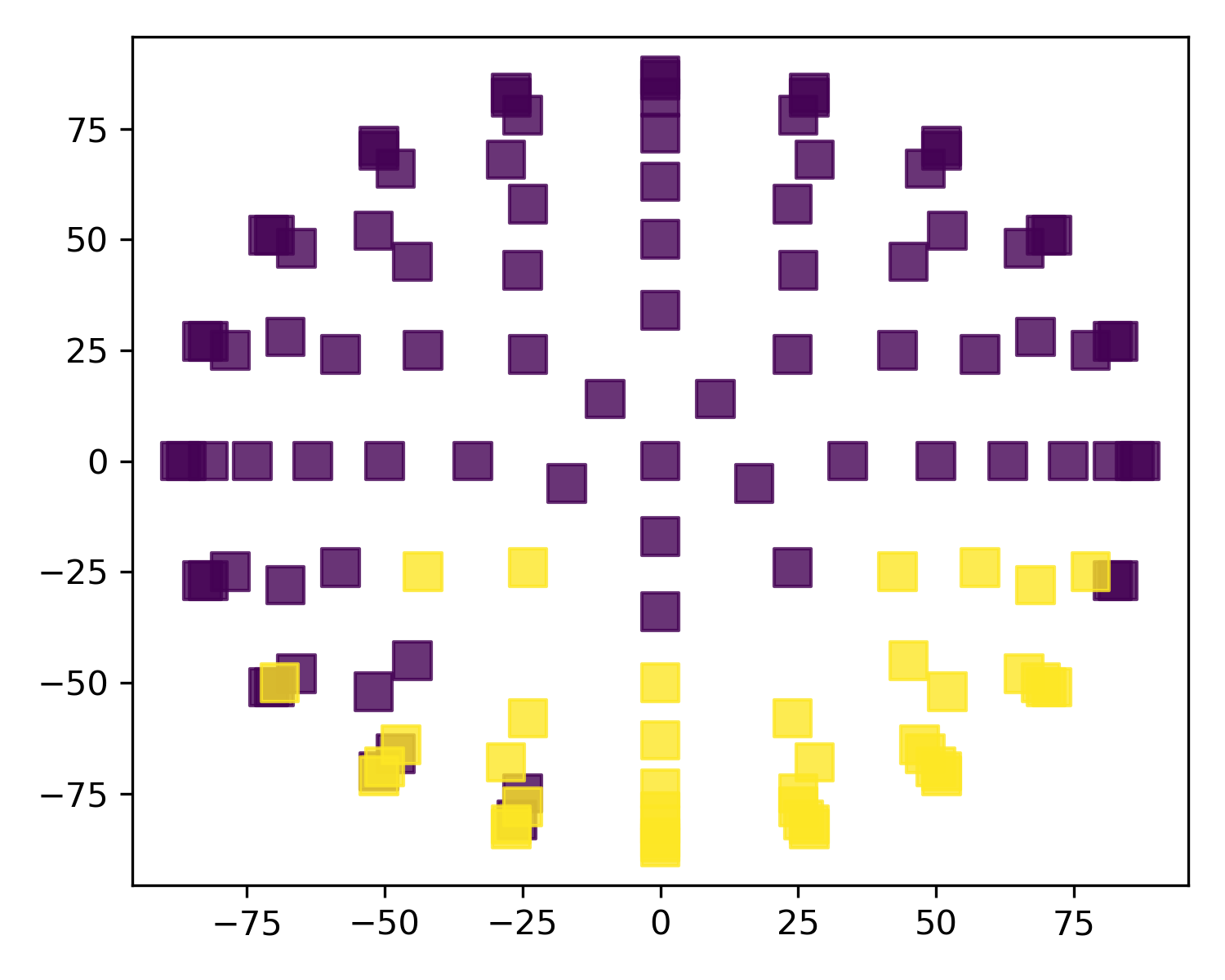}
                             \caption{\small The subject has two communities, with the larger one near the front of the head.}
  \end{subfigure}
 \begin{subfigure}[t]{0.32\textwidth}
        \includegraphics[width=1\linewidth, height=2in]{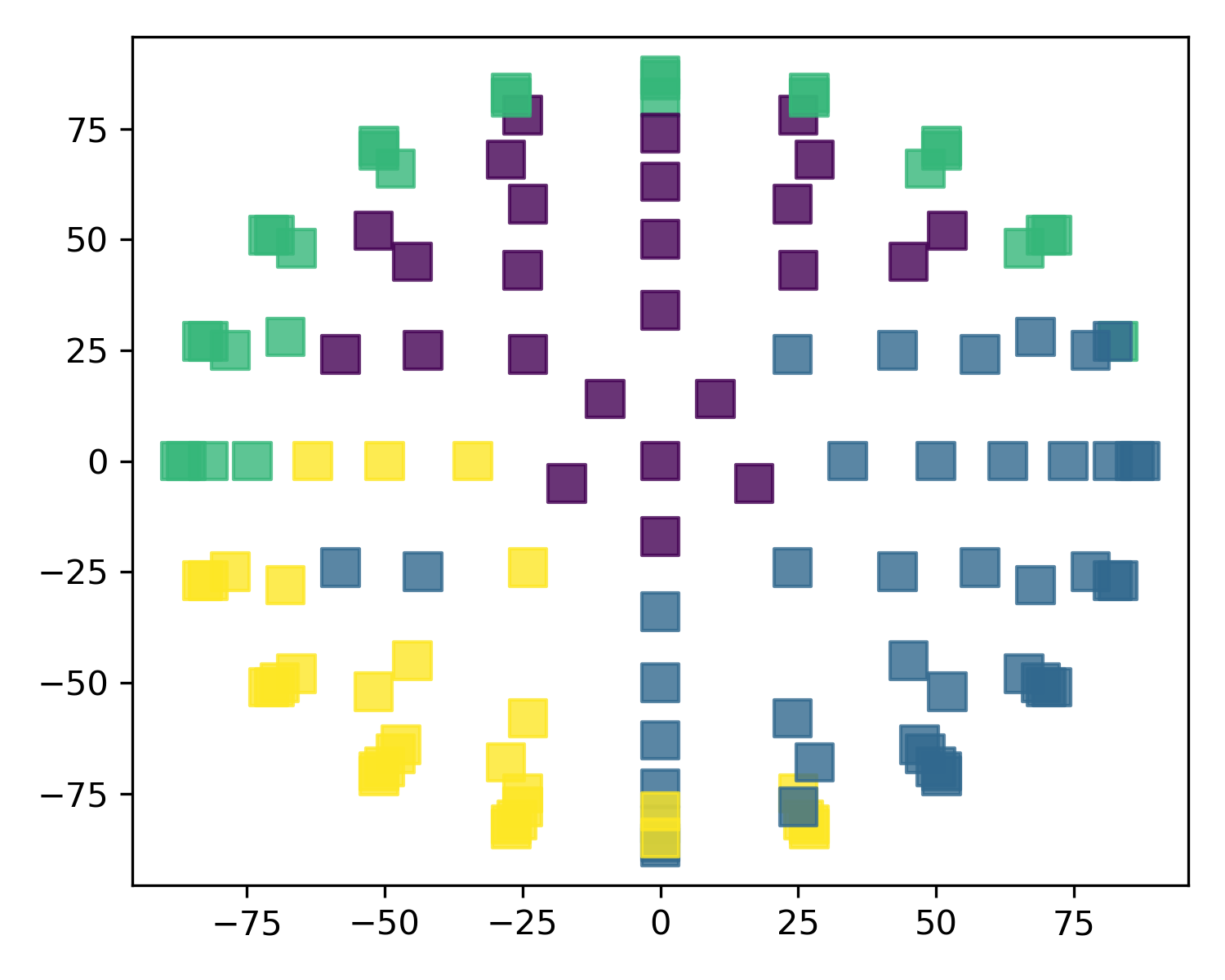}
                                    \caption{\small The subject has four communities: outer-front, mid-front, left-back, right-back.}
  \end{subfigure}
   \caption{Community structure for each brain scan from multiple subjects in the working memory study.\label{fig:com_structure}}
  \end{figure}

 Figure ~\ref{fig:working_memory_data} depicts the adjacency matrices of three subjects for the memory retrieval task,
 and the presence of heterogeneity is apparent. It can be seen that memory-related connectivity can exhibit different levels of concentration in the front or back of the head [Panels (a) or (b), with spatial coordinates, plotted in Figure~\ref{fig:com_structure} (a)], or, they are more localized in smaller regions [Panel (c)].
 
We apply the spiked graph Laplacian model on this data set and the results obtained are based on an MCMC run of
 $30,000$ steps, with the first $10,000$ used as the burn-in period.
 The majority of the samples from the posterior distribution contain six distinct $U^{(l)}$'s in the clustered eigenmatrix values. Figure ~\ref{fig:fitted-Laplacian} depicts the three corresponding to the raw $A^{(s)}$ shown in the previous Figure, obtained from the fitted Laplacian matrices. The remaining three seem to correspond to smaller variations and are shown in the supplementary materials. The proportions for these six groups are $25.6\%,24.1\%,16.1\%, 14.7\%, 15.2\%$ and $4.3\%$, as estimated in the posterior mean of allocation $z_s$.

 We then evaluate the community structures in each network.   As shown in Figure~\ref{fig:com_structure}(a), the model discovers $1\sim 6$ communities from these graphs, as estimated by $\kappa^{(s)}$. To gain insight into the scientific implications, we plot the community labels mapped to the spatial coordinates.  Panel (c) and (d) show that most of the networks contain only two distinct communities, although the division can be quite different in the dominating area either in the front or in the back of the head. Panel (e) shows a very distinct pattern with four communities, partitioned as the outer-front, mid-front, left-back, right-back regions of the head.
  
\section{Discussion}
In this paper, we propose a probabilistic graph model based on the Laplacian, allowing us to exploit the spectral graph theory to conduct flexible community detection in a population of heterogeneous graphs. Our model can be considered as a general idea to introduce Bayesian toolboxes into the spectral graph framework. There are several extensions worth exploring in future work. First, if the goal is to generate a new graph with binary $A_{i,j}$, such as in link prediction, then it could adopt a Bernoulli distribution associated with a canonical link.  Second, if those graphs have some known covariance structure, such as repeated measurement or temporal effect,  then it could take an alternative distribution on the eigenmatrix or eigenvalues to incorporate those structures. Lastly, for large graphs, it is of interest to consider 
$\theta$ not as a single constant, but as a step function.

\spacingset{0}

\bibliography{reference}
\bibliographystyle{chicago}

\subsection*{Supplementary Materials}

{\footnotesize
\subsection*{\small Proof of Lemma 1}
\begin{proof}
The bounds on the eigenvalues of the Laplacian are given and discussed in  \cite{chung1997spectral}. For the first eigenvector we have
\be
 \mu_L \vec d_*^{1/2}= 
D_*^{-1/2} (D_*-A_*)D_*^{-1/2} \vec d_*^{1/2}
=D_*^{-1/2} (D_*-A_*) \vec 1 =
\vec 0.
 \ee
\end{proof}

\subsection*{\small Proof of Theorem 1}

\begin{proof}
For simplicity, we omit $.^{(s)}$ in the proof and use $\sigma_e$ for $\sigma_{e0}$. The proof consists of the following four parts:

    {\bf 1. An application of the Davis-Kahan Theorem}

Let $E= \tilde L - L$, using Theorem 2 in citep{yu2014useful} with $r=1$ and $s=k$, we obtain
    \be
       &  \| Q_0 -  Q O\|_F  \le \frac{2^{3/2} \min(k^{1/2} \|E\|_{op},\|E\|_F)}{ \lambda_{k+1} - \lambda_k} \\
      &  \le \frac{2^{3/2} (k^{1/2} \|E\|_{op})}{\lambda_{k+1} - \lambda_k}
    \ee
    where $\|E\|_{op}$ denotes the operator norm ($\|E\|_{op} = \sup_{\|x\|=1} \|E x\|$).

        {\bf 2. Discretizing $\mathbb{S}^{n-1} = \{x: \|x\|=1\}$ using a maximal $\epsilon$-net}:

        Following \cite{tao2012topics}, let $N_\epsilon\subset \mathbb{S}^{n-1} $ be an $\epsilon$-net with $\epsilon\in(0,1)$, such that for any two $x\in \mathcal{N}_\epsilon, x'\in \mathcal{N}_\epsilon, \|x-x'\|\ge\epsilon$. Maximizing over the number of included points in $\mathbb{S}^{n-1}$, we obtain a maximal $\epsilon$-net $N_\epsilon^0$. Clearly, the balls with centers $x\in N_\epsilon^0$ and radius $\epsilon/2$ are disjoint, and all covered by a large ball centered at the origin with radius $1+\epsilon/2$, hence
        \be
        |\mathcal{N}^0_\epsilon| \le (\frac{\epsilon/2+1}{\epsilon/2})^n =  (\frac{\epsilon+2}{\epsilon})^n.
        \ee
        On the other hand, for any $y\in \mathbb{S}^n$, there is at least one $x\in \mathcal{N}_\epsilon^0: \|x-y\|\le \epsilon$, otherwise $y$ can be added to the net, contradicting the maximal condition.

        Choosing $y\in \mathbb{S}^n$ that attains $\|E y \| = \|E\|_{op}$, and its associated $x\in \mathcal{N}_\epsilon^0: \|x-y\|\le \epsilon$
            \be
       \|E\|_{op} - \|E x\|   = \|E y\| - \|E x\| \le \|E (y- x)\|  \le \|E\|_{op} \epsilon,
        \ee
       by an application of the triangle inequality and $f(x)=\|Ex\|$ is $\|E\|_{op}$-Lipschitz.

        Therefore, $\|E\|_{op}\ge t$ implies at least one $x\in \mathcal{N}_\epsilon^0: \|E x\| \ge (1-\epsilon) t$.
        \be
        \text{pr}(\|E\|_{op}\ge t)
        & \le  \text{pr}\bigg(\bigcup_{x\in \mathcal{N}_\epsilon^0}  \|E x\| \ge (1-\epsilon) t \bigg) \\
        & \le  |\mathcal{N}^0_\epsilon|  \text{pr}\bigg(  \|E x\| \ge (1-\epsilon) t   \text{, where } x\in \mathbb{S}^n \bigg)
        \ee
       where the last inequality follows from the union bound.

     {\bf 3. Concentration inequality for $\|Ex\|$}
     
Since $E$ is symmetric, let $E = E_U + E_L$, with $E_U$ being the upper triangular portion including the diagonal and $E_L$ the lower triangular portion.  We first use $B$ to represent either $E_U$ or $E_L$. 
     Let $B$ be an $n\times n $ matrix comprising of $b_{i,j}$ independent and $\sigma^2_e$-sub-Gaussian elements. Then, for each element $Bx$
     \be
     \mathbb{E} \exp \{t  B_j^{\prime} x \}
     & = \mathbb{E} \exp \{t  \sum_{k=1}^n x_k b_{j,k} \} \\
   & = \prod_{k=1}^n\mathbb{E} \exp \{t   x_k b_{j,k} \} \\
      & \le \prod_{k=1}^n  \exp \{t^2 \sigma^2_e  x^2_k /2 \}\\
      & =  \exp \{t^2 \sigma^2_e   /2 \}
      \ee
      where the inequality is due to the sub-Gaussian assumption, and the last equality due to $\|x\|=1$. Therefore, each $Z_j=B_jx$ is sub-Gaussian as well. 
      By a result in \cite{wainwright2017highdim}, this is equivalent to
      \bel{eq:sub_gaussian_quad}
      \mathbb{E} \exp( \frac{\kappa Z_j^2}{2\sigma^2_e} ) \le (1-\kappa)^{-1/2}
      \eel
      for all $\kappa\in(0,1)$.

        We have
      \be
         \|Ex\|^2 = \|E_Ux + E_L x\|^2 \le (\|E_Ux\| + \|E_L x\|)^2 \le 2(\|E_Ux\|^2 + \|E_L x\|^2)
      \ee
              By the Cauchy--Schwarz inequality, we obtain
           \be
      &\mathbb{E} \exp( \frac{\kappa  \|Ex\|^2 }{2\sigma^2_e} )
           \le        \mathbb{E} \exp( \frac{2  \kappa  [ \|E_U x\|^2 + \|E_L x\|^2] }{2\sigma^2_e} )  \\
      &\le
      \sqrt{     \mathbb{E} \exp( \frac{4\kappa  \|E_U x\|^2 }{2\sigma^2_e} )  \mathbb{E}\exp( \frac{4\kappa  \|E_Lx\|^2 }{2\sigma^2_e} ) }
            \ee
            Since $E_U$ and $E_L$ comprise of sub-Gaussian elements and zeros, they are also sub-Gaussian with $\sigma^2_e$; then, 
            multiplying \eqref{eq:sub_gaussian_quad} over $j=1,\ldots,n$ for each matrix, we get

      \be
      \mathbb{E} \exp( \frac{\kappa  \|Ex\|^2 }{2\sigma^2_e} ) \le \sqrt{(1- 4\kappa)^{-n/2}  (1- 4\kappa)^{-n/2} } = (1- 4\kappa)^{-n/2}
      \ee
      where $\kappa \in(0,1/4)$.
        Using Markov's inequality
        \be
        \text{pr}(\|Ex\| \ge t) = \text{pr} \bigg(
        \exp( \frac{\kappa  \|Ex\|^2 }{2\sigma^2_e} )  \ge
        \exp( \frac{\kappa  t^2 }{2\sigma^2_e} )
        \bigg)
        \le (1-4\kappa)^{-n/2}  \exp( -\frac{\kappa  t^2 }{2\sigma^2_e} ) .
        \ee

          {\bf 4. Combining results to obtain a concentration inequality}

        Therefore,

\be
\text{pr}(\|E\|_{op}\ge t)  \le
(\frac{\epsilon+2}{\epsilon})^n
 (1-4\kappa)^{-n/2}     \exp( -\frac{\kappa  (1-\epsilon )^2 t^2 }{2\sigma^2_e} )
\ee
Letting $t=c_1 \sqrt{n} \sigma_e $, $\kappa = 1/8$ and $\epsilon = 1/2$, we have
\be
\text{pr}(\|E\|_{op}\ge c_1 \sqrt{n} \sigma_e)  \le
\exp[ -  \{c_1^2/64 - \log(5 \sqrt{2})\} n ] \equiv \delta
\ee
Therefore,
\be
    \| Q - \hat Q \hat O\|_F \le \frac{2^{3/2} k^{1/2} c_1 \sqrt{n} \sigma_e}{ \lambda_{k+1} - \lambda_k}
    \ee
    with probability greater than $1-\delta$.

\end{proof}

\vspace{-8mm}
\subsection*{\small Proof of Theorem 2}
\vspace{-3mm}

\begin{proof}

For simplicity, we omit $.^{(s)}$ for now and let $D = \Lambda^\dagger$ and $B = \Omega$.
Without loss of generality, we assume the diagonal of $B$ are ordered $0 = b_1 < b_2 \le \ldots \le b_n$; and we have fixed $d_1=0$ and $d_2,\ldots,d_n>0$. The parameter $Q^\dagger$ follows a matrix Bingham distribution truncated to $\mathcal V^*$
\be
\tilde g(Q^\dagger; W, D, B, \sigmae) \Pi (d Q^\dagger )=Z^{-1} (\sigmae, D,B)
\etr\bigg\{
         \frac{1}{2\sigma_e^2}  D Q ^{\dagger \prime} W B W ^{\prime} Q^\dagger   \bigg\}
\Pi ( \textup{d} Q^\dagger)
\ee
where $Z$ is a normalizing constant.

We utilize the result of \cite{bhattacharya2010nonparametric} to establish weak consistency of the posterior density estimation. There are three sufficient conditions to check:

(1) The kernel $\tilde g(.)$ is continuous in all of its arguments.

(2) The set  $\{F_0\} \times D^0_\epsilon$ intersects the parameter support of $Q^\dagger$ and $\sigmae$, where $\mathcal D^0_\epsilon$ is the interior of  $\mathcal D_\epsilon$, a compact neighborhood for $\sigmae$.

(3) For any  continuous $f$, there is a $\mathcal D_{\epsilon}$ for $\sigmae$, such that
\be
\Delta =\sup_{W \in \mathcal V^*,\sigmae \in \mathcal D_\epsilon}
\bigg\|
f(W) - \int  \tilde g(Q^\dagger; W, D, B, \sigmae)  f(Q^\dagger) \Pi( \textup{d}Q^\dagger)
\bigg\|
\le \epsilon.
\ee

The first two conditions are straightforward to check (see  \cite{lin2017bayesian} for similar derivation). We will focus on verifying (3). Note the Frobenius distance between two orthonormal matrices
\be
\text{dist}(W,Q^\dagger)^2 =  2 n  - 2\tr( W^{\prime}Q^\dagger) = 2 \sum_{j=1}^n (1- g_{j,j}),
\ee
where $g_{i,j}$ is the element of $G= W ^{\prime}Q^\dagger $, where $|g_{j,j}|\le 1$ due to orthonormality of $G$. Let $  (1- g_{j,j})= s_{j,j} \sigma_e$, with $s_{j,j}\in[0, 2/\sigma_e]$, then $\sum_{j=1}^n (1- g_{j,j}) = \sum_{j=1}^n s_{j,j} \sigma_e$. As $\sigma_e \to 0$, $\text{dist}(W,Q^\dagger) \to 0$ for any fixed $(s_{1,1},\ldots, s_{n,n})$. By the continuity of $f$ and compactness of Stiefel manifold, as $\sigma_e \to 0$
\bel{eq:proof_dist1}
\sup_{W \in \mathcal V^*}
\bigg\|
f(W) -f(Q^\dagger)\bigg\|
\to 0.
\eel

Now
\be
\Delta& \le Z^{-1} (\sigmae, D,B)
\int
 \sup_{W \in \mathcal V^*}
\bigg\|
f(W) - f(Q^\dagger)\bigg\|
 \etr  \bigg\{ \frac{1}{2\sigmae } D Q ^{* \prime} [W BW ^{\prime}] Q^\dagger
\bigg\}
 \Pi( \textup{d}Q^\dagger)
 \\
& = Z^{-1} (\sigmae, D,B)
\int
 \sup_{W \in \mathcal V^*}
\bigg\|
f(W) - f(WG)\bigg\|
 \etr  \bigg\{ \frac{1}{2\sigmae}  D G ^{\prime} B G
\bigg\}
\Pi( \textup{d}G)
 \\
\ee
where the second line is due to the invariant volume of rotation via $W$. It can be verified that
\be
\tr &( D G ^{\prime} B G) =  \sum_{i=1}^n    \sum_{j=1}^n b_{i} d_{j} g_{i,j}^2 \\
& =  \sum_{j=1}^n b_j d_j -  \sum_{j=1}^n b_j d_j (1- g_{j,j}^2) +  \sum_{j=1}^n \sum_{i\ne j} b_{i} d_{j}  g_{i,j}^2\\
& \le   \sum_{j=1}^n b_j d_j -  \sum_{j=1}^n b_j d_j (1- g_{j,j}^2) +  \sum_{j=1}^n d_{j}  b_n \sum_{i\ne j}   g_{i,j}^2 \\
& =  \sum_{j=1}^n b_j d_j -  \sum_{j=1}^n b_j d_j (1- g_{j,j}^2) +  \sum_{j=1}^n d_{j}  b_n   (1- g_{j,j}^2) \\
& =  \sum_{j=1}^n b_j d_j + \sum_{j=1}^n d_{j}  (b_n -b_j) (1- g_{j,j}^2) \\
& =  \sum_{j=1}^n b_n d_j  - \sum_{j=1}^n d_{j}  (b_n -b_j)  g_{j,j}^2 ,
\ee
where the first inequality is due to $d_j \ge 0$ and $b_n\ge b_i$ for all $i$; the
fourth line is due to the $1$ unit norm for each column of $G$.

Applying one-to-one transformation $T:\mathcal V^*\to \mathcal S$, $T(G)= \{ s_{i,j} = g_{i,j} \text{ for } i\neq j, s_{j,j}=(1-g_{j,j}) /\sigma_e \}_{i,j}$,  denote the transformed $G$ matrix by $G_S$. We have
 \be
& \Pi( dG)= \phi(G)  \text{d}g_{1,1} \wedge \text{d} g_{1,2}\wedge\ldots \wedge \text{d} g_{n,n}\\ 
& =\frac{\phi(G)} {\tilde\phi(G_S)} \sigma^{n}_e  \tilde \phi(G_S)\text{d}s_{1,1} \wedge \text{d} s_{1,2}\wedge\ldots
\wedge \text{d} s_{n,n} \\
& =\frac{\phi(G)} {\tilde\phi(G_S)}  \sigma^n_e  \Pi( \textup{d}G_{S}),
 \ee
 where $\phi$ and $\tilde\phi$ are some functions of $G$ and $G_s$, respectively.

Since $s_{j,j}\le 2/\sigma_e$, we have $ -(1-s_{j,j}\sigma_e)^2=  -1+ 2s_{j,j}\sigma_e - s^2_{j,j}\sigma^2_e \le
 3- s^2_{j,j}\sigma^2_e$. Continuing from above,

 \be
  &   \sum_{j=1}^n b_n d_j -     \sum_{j=1}^n d_{j}  (b_n -b_j)
 (1-s_{j,j}\sigma_e)^2 \\
 &\le
  \sum_{j=1}^n b_n d_j +     \sum_{j=1}^n
d_{j}  (b_n -b_j)
 (3- s^2_{j,j}\sigma^2_e)
\\
& =
  \sum_{j=1}^n 4b_n d_j -     \sum_{j=1}^n
3d_{j}  b_j
 -     \sum_{j=1}^n
d_{j}  (b_n -b_j)
  s^2_{j,j}\sigma^2_e\\
 & \le \sum_{j=1}^n 4b_n d_j 
 -     \sum_{j=1}^n
d_{j}  (b_n -b_j)
  s^2_{j,j}\sigma^2_e
 \ee
 
 Combining the above,
\bel{eq:bounded_terms}
\Delta \le  Z^{-1} &(\sigmae, D,B)
  \exp  \bigg[ \frac{1}{2\sigmae}
  (
  \sum_{j=1}^n 4b_n d_j -     \sum_{j=1}^n
3d_{j}  b_j
 )
   \bigg] 
   \sigma_e^n
   \int_{\mathcal S}
 \sup_{W \in \mathcal V^*}
\bigg\|
f(W) - f(WG_S)\bigg\| \\
& \times\exp \bigg[-    \frac{1}{2  }        \sum_{j=1}^n
d_{j}  (b_n -b_j)
  s^2_{j,j}
 \bigg]
\frac{\phi(G)} {\tilde\phi(G_S)}   \Pi( \textup{d}G_{S}).
\eel

Note that $\sup_{G_S\in \mathcal V^*} \sup_{W \in \mathcal V^*}
\bigg\|
f(W) - f(WG_S)\bigg\|\le M$ due to the compactness of $\mathcal V^*$ and continuity of $f$. And clearly,

\be
   \int_{\mathcal S}
M\exp \bigg[-    \frac{1}{2  }        \sum_{j=1}^n
d_{j}  (b_n -b_j)
  s^2_{j,j}
 \bigg]
\frac{\phi(G)} {\tilde\phi(G_S)}   \Pi( \textup{d}G_{S}) <\infty.
 \ee
 Using dominated convergence theorem, when $\sigma_e \to 0$, the integral in \eqref{eq:bounded_terms} goes to zero.

 Our remaining task is to verify the constant before the integral is finite as $\sigma_e \to 0$. Note the inverse of the constant in \eqref{eq:bounded_terms}
\bel{eq:const}
\sigma_e^{-n}Z&(\sigmae, D,B)
  \exp  \bigg[- \frac{1}{2\sigmae}
   \sum_{j=1}^n 4b_n d_j   \bigg] \\
&=\sigma_e^{-n}
  \exp  \bigg[- \frac{1}{2\sigmae}
   \sum_{j=1}^n 4b_n d_j   \bigg]
\int_{ \mathcal V^*}
 \etr  \bigg\{ \frac{1}{2\sigmae} D U ^{\prime}  B U  \bigg\}
  \Pi(\textup{d}U) \\
 &=\sigma_e^{-n}
\int_{ \mathcal V^*}
 \exp  \bigg\{ \frac{1}{2\sigmae}  (\sum_{i=1}^n    \sum_{j=1}^n b_{i} d_{j} u_{i,j}^2 -  \sum_{j=1}^n 4b_n d_j  \sum_{i=1}^n u^2_{i,j} ) \bigg\}
 \Pi(\textup{d}U)\\
    &=\sigma_e^{-n}
\int_{ \mathcal V^*}
 \exp  \bigg\{ \frac{1}{2\sigmae} \sum_{i=1}^n    \sum_{j=1}^n (b_{i} -4b_n)d_{j} u_{i,j}^2  \bigg\}
  \Pi(\textup{d}U)\\
      &=\sigma_e^{-n}
\int_{ \mathcal V^*}
 \exp  \bigg\{ \frac{1}{2\sigmae} \sum_{i=1}^n    \sum_{j=2}^n (b_{i} -4b_n)d_{j}
u_{i,j}^2  \bigg\}
  \Pi(\textup{d}U)\\
        &\ge\sigma_e^{-n}
\int_{ \mathcal V^*}
 \exp  \bigg\{ \frac{1}{2\sigmae} \sum_{i=1}^n    \sum_{j=2}^n (b_{i} -4b_n)d_{n}
u_{i,j}^2  \bigg\}
  \Pi(\textup{d}U)\\
          &=\sigma_e^{-n}
\int_{ \mathcal V^*}
 \exp  \bigg\{ \frac{1}{2\sigmae} \sum_{i=1}^n    (b_{i} -4b_n)d_{n}
(1-u_{i,1}^2)  \bigg\}
  \Pi(\textup{d}U),
  \eel
 where we  use $d_1=0$, $(b_{i} -4b_n)\le 0$ and $d_j\le d_n$ in the inequality.
Since the last line does not depend on $U_{2:n}$, we denote the null space of $u_1$ by $\mathcal K(u_1)=\{ U_{2:n}\in \mathcal V^{n-1,n}: u_k'u_1=0, k>1 \}$. It is not hard to see that the volume $\mathcal K(u_1)$ is a constant invariant to $u_1$, we denote it by $vol(\mathcal K)$. The above is then,
\be
& \sigma_e^{-n}vol(\mathcal K)
\int_{ \mathbb S_+}
 \exp  \bigg\{ -\frac{1}{2\sigmae} \sum_{i=1}^n    (4b_n-b_i)d_{n}
(1-u_{i,1}^2)  \bigg\}
  \Pi(\textup{d}U_1) \\
  &\ge \sigma_e^{-n}vol(\mathcal K)
\int_{ \mathbb S_+}
 \exp  \bigg\{ -\frac{1}{2\sigmae} \sum_{i=1}^n    (4b_n-b_i)d_{n}
(1+u_{i,1})^2  \bigg\}
 \Pi( \textup{d}U_1 ),\\
\ee
where $\mathbb{S}_+$ is the unit-norm space constrained to all elements positive; and the inequality due to $-(1-u^2)=-(1-u)(1+u)\ge -(1+u)^2$ for $u\ge0$.

Let $t_{i}=(1+u_{i,1})/\sigma_e$. We have
 \be
& \Pi( \textup{d}U_1 )= \psi(U_1)  \text{d}u_{1,1} \wedge \text{d} u_{2,1}\wedge\ldots
\wedge \text{d} u_{n,1}\\ 
& =\frac{\psi(U_1)} {\tilde\psi(T)} \sigma^{n}_e  \tilde \psi(T)\text{d}t_{1}
\wedge \text{d}t_2\wedge\ldots
\wedge \text{d}t_n \\
& =\frac{\psi(U_1)} {\tilde\psi(T)}  \sigma^n_e  \Pi(\textup{d} T),
\ee
where $\psi$ and $\tilde\psi$ are some functions of $U_1$ and $T$, respectively.

The above is then 
\be
vol(\mathcal K)
\int_{\mathcal T}
 \exp  \bigg\{ -\frac{1}{2} \sum_{i=1}^n    (4b_n-b_i)d_{n}
t_{i}^2  \bigg\}
 \frac{\psi(U_1)} {\tilde\psi(T)}  \Pi(\textup{d} T),
\ee
which is bounded away from $0$. Therefore, the constant in  \eqref{eq:bounded_terms} is finite as $\sigmae \to 0$.

The limit result means that for any $\epsilon>0$, we have a neighborhood $\mathcal D_\epsilon = \{\sigmae: 1/\sigmae> N_\epsilon\}$, so that $\Delta<\epsilon$.

\end{proof}

\subsection*{Details of the Gibbs Sampling Algorithm} \label{Gibbs-algo}
The posterior sampling proceeds according to the following steps:

\begin{enumerate}
    \item Sample $R_s$  from (11) in the main article.
    \item Sample $U^{(l)}$ from (12) in the main article.
  \item Sample from the categorical distribution
    \be
    z_s \sim \Pi(z_s \mid .) \propto \pi_l  & 1(z_s=l )\exp\bigg\{
\frac{1}{2}
(\frac{n-T}{2\sigmae} + \frac{1}{\sigma^2_\theta})^{-1}
 \bigg[
 \frac{1}{2\sigmae}  \tr (\big[ L^{(s)} (I_n - U^{(l)} U^{(l)T})
\big]) + \frac{\mu_\theta}{\sigma^2_\theta}
\bigg ]^2 \\
& +
\frac{1}{2}
(\frac{1}{\sigma^2_{\lambda,\eta^{(s)}_k}} +\frac{1}{2\sigma^2_e})^{-1} \sum_{k=1}^T
\bigg[ \frac{ u_k^{(l)\prime}L^{(s)}  u^{(l)}_k}{2\sigmae}
+ \frac{(1-\eta^{(s)}_k) \mu_\theta}{\sigma^2_{\lambda,\eta^{(s)}_k}}\bigg]^2  \bigg\},
    \ee      with $1(.)$ the indicator function, update $Q^{(s)} = U^{(z_s)}.$
    \item Sample $(\pi_1,\pi_2,\ldots, \pi_g) \sim \text{Dir}(\alpha_0/g
+ \sum 1(z_s =1) ,\alpha_0/g + \sum 1(z_s =2) ,\ldots, \alpha_0/g+
 \sum 1(z_s =1))$.
    \item Sample for $k = 2,\ldots, T$
    \be
    \lambda^{(s)}_k \sim \No_{(0,2)}\bigg\{ (\frac{1}{\sigma^2_{\lambda,\eta^{(s)}_k}} +\frac{1}{2\sigma^2_e})^{-1} \bigg[ \frac{ q_k^{(s)\prime}L^{(s)}  q^{(s)}_k}{2\sigmae} + \frac{ (1-\eta^{(s)}_k)\mu_\theta}{\sigma^2_{\lambda,\eta^{(s)}_k}}\bigg] , (\frac{1}{\sigma^2_{\lambda,\eta^{(s)}_k}} +\frac{1}{2\sigma^2_e})^{-1}  \bigg\}.
\ee
    \item Sample from the Bernoulli for $k = 2,\ldots, T$,
    \be
   \eta^{(s)}_k\sim &   1(\eta^{(s)}_k=1)w
    \No_{(0,2)} (\lambda_k^{(s)}; 0,\sigma^2_{\lambda,1} ) +
 1(\eta^{(s)}_k=0)(1-w)
    \No_{(0,2)} (\lambda_k^{(s)} ; \mu_\theta,\sigma^2_{\lambda,0} ) ,
    \ee
    where $\No_{(0,2)} (x; a,b)$ denotes the density of the truncated normal. 
\item Sample
\be
\theta^{(s)} \sim \No_{(0,2)}\bigg\{ &
(\frac{n-T}{2\sigmae} + \frac{1}{\sigma^2_\theta})^{-1}
\bigg[
 \frac{1}{2\sigmae} (\sum_i L^{(s)}{ (i,i)}  -  \sum_k q_k^{{(s)}\rm{T}} L^{(s)} q^{(s)}_k) + \frac{\mu_\theta}{\sigma^2_\theta}
\bigg]
,\\ &
(\frac{n-T}{2\sigmae} + \frac{1}{\sigma^2_\theta})^{-1}
\bigg\}.
\ee
\item Sample for $i=1,\ldots,n$
\be
L^{(s)}_{i,i} \sim \No \bigg\{  \big[ Q^{(s)} (\Lambda^{(s)} -\theta^{(s)} I_T) Q^{{(s)}\prime} \big]_{(i,i)} + \theta^{(s)},2\sigma^2_e \bigg\}.
\ee
\item Sample
\be
\sigma^2_e \sim \text{Inv-Gamma} \bigg\{  \frac{n^2S}{2},\frac{1}{4}
 \sum_{s=1}^S\| L^{(s)} - \theta  I_n - Q_*^{(l)} (\Lambda^{(s)} -\theta^{(s)} I_T) Q_*^{{(l)}\prime} \|_F^2  \bigg\}.
\ee
\end{enumerate}
}

\subsection*{Simulation for Estimating Latent Structure}

\begin{figure}[H]
 \begin{subfigure}[t]{0.22\textwidth}
 \centering
       \includegraphics[width=1\linewidth]{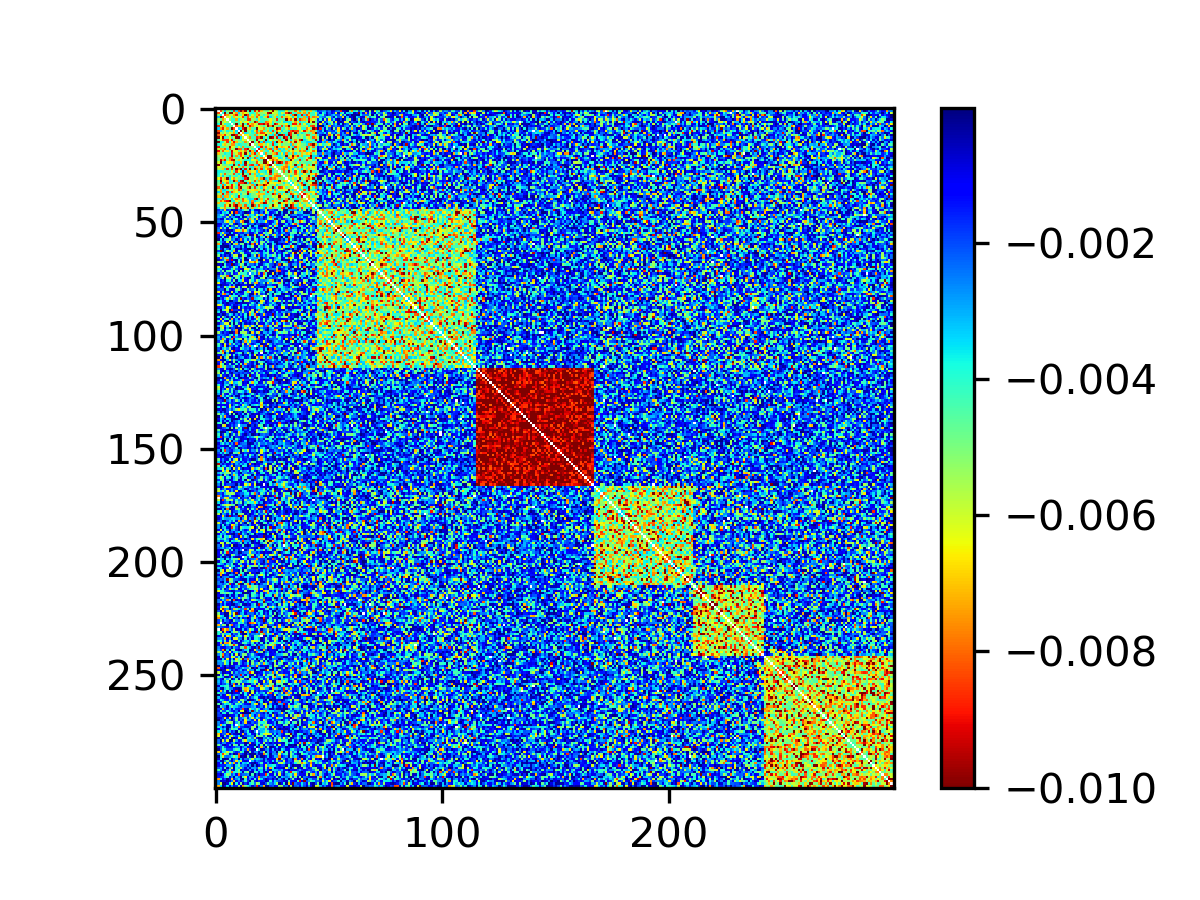}
       \caption{Laplacian of a simulated graph.}
   \end{subfigure}
 \begin{subfigure}[t]{0.22\textwidth}
 \centering
       \includegraphics[width=1\linewidth]{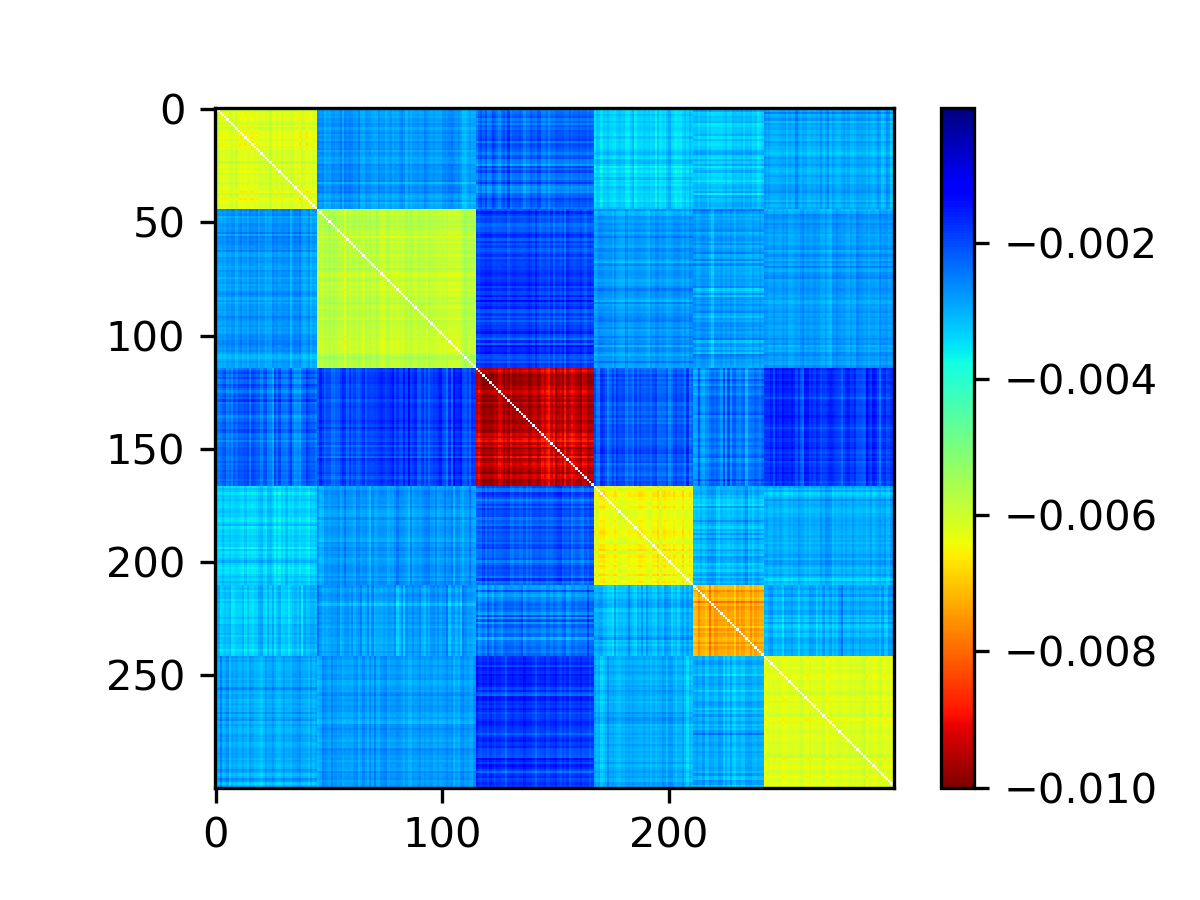}
       \caption{Estimated spiked Laplacian.}
 \end{subfigure}
  \begin{subfigure}[t]{0.22\textwidth}
 \centering
       \includegraphics[width=1\linewidth]{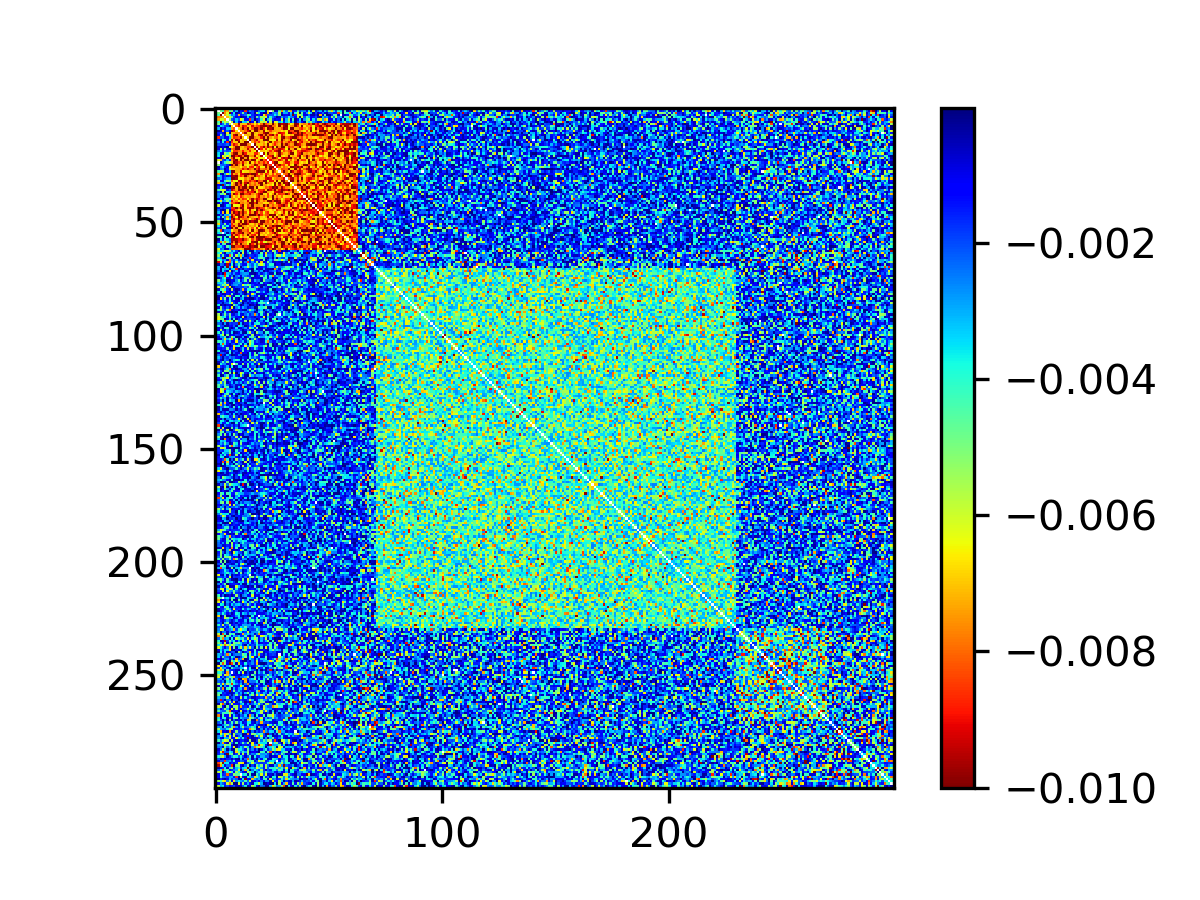}
       \caption{Laplacian of a simulated graph.}
   \end{subfigure}
     \begin{subfigure}[t]{0.22\textwidth}
  \centering
       \includegraphics[width=1\linewidth]{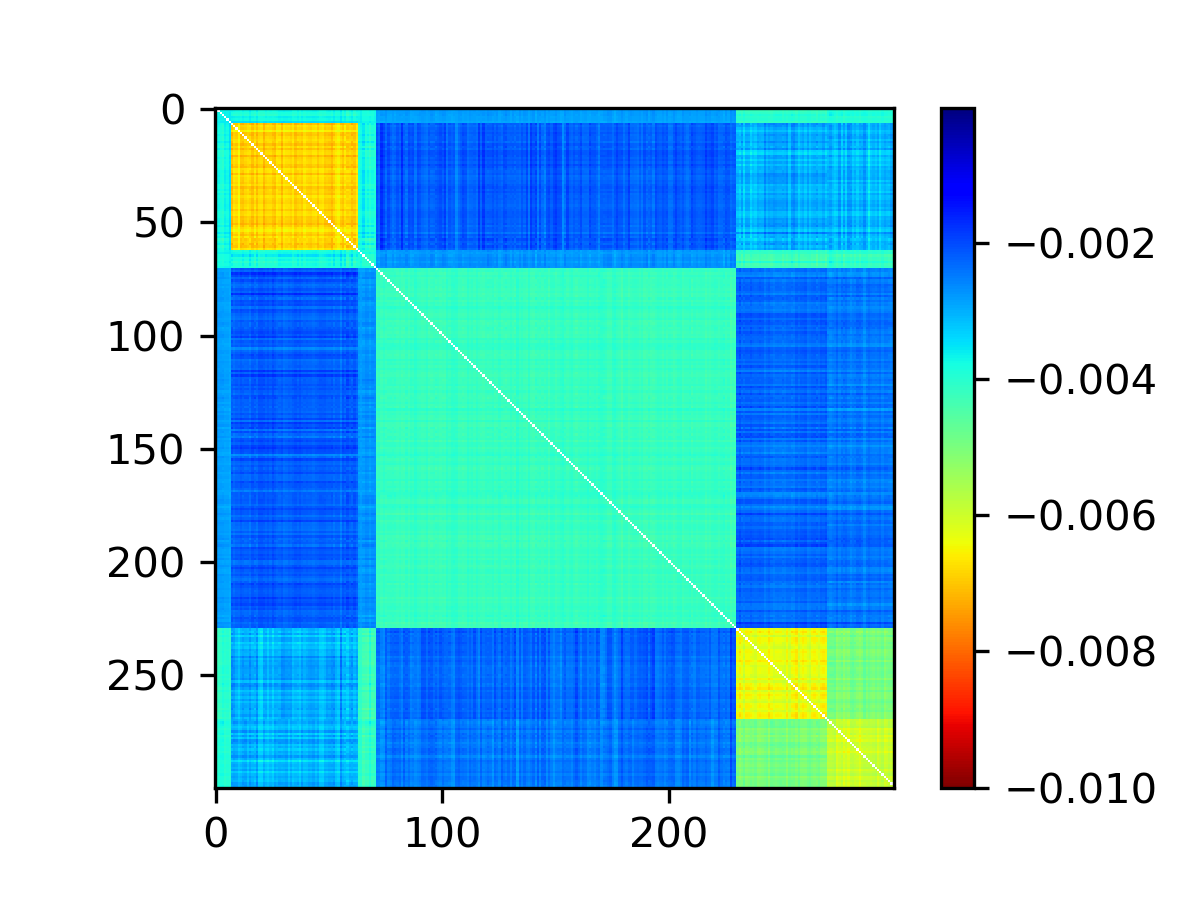}
       \caption{Estimated spiked Laplacian.}
   \end{subfigure}
 \caption{The proposed model correctly finds the latent community structures in $200$ simulated graphs. \label{fig:sim_heterogeneity}}
 \end{figure}

\subsection*{Additional Components in Working Memory Data Analysis}

\begin{figure}[H]
 \begin{subfigure}[t]{0.32\textwidth}
 \centering
       \includegraphics[width=1\linewidth]{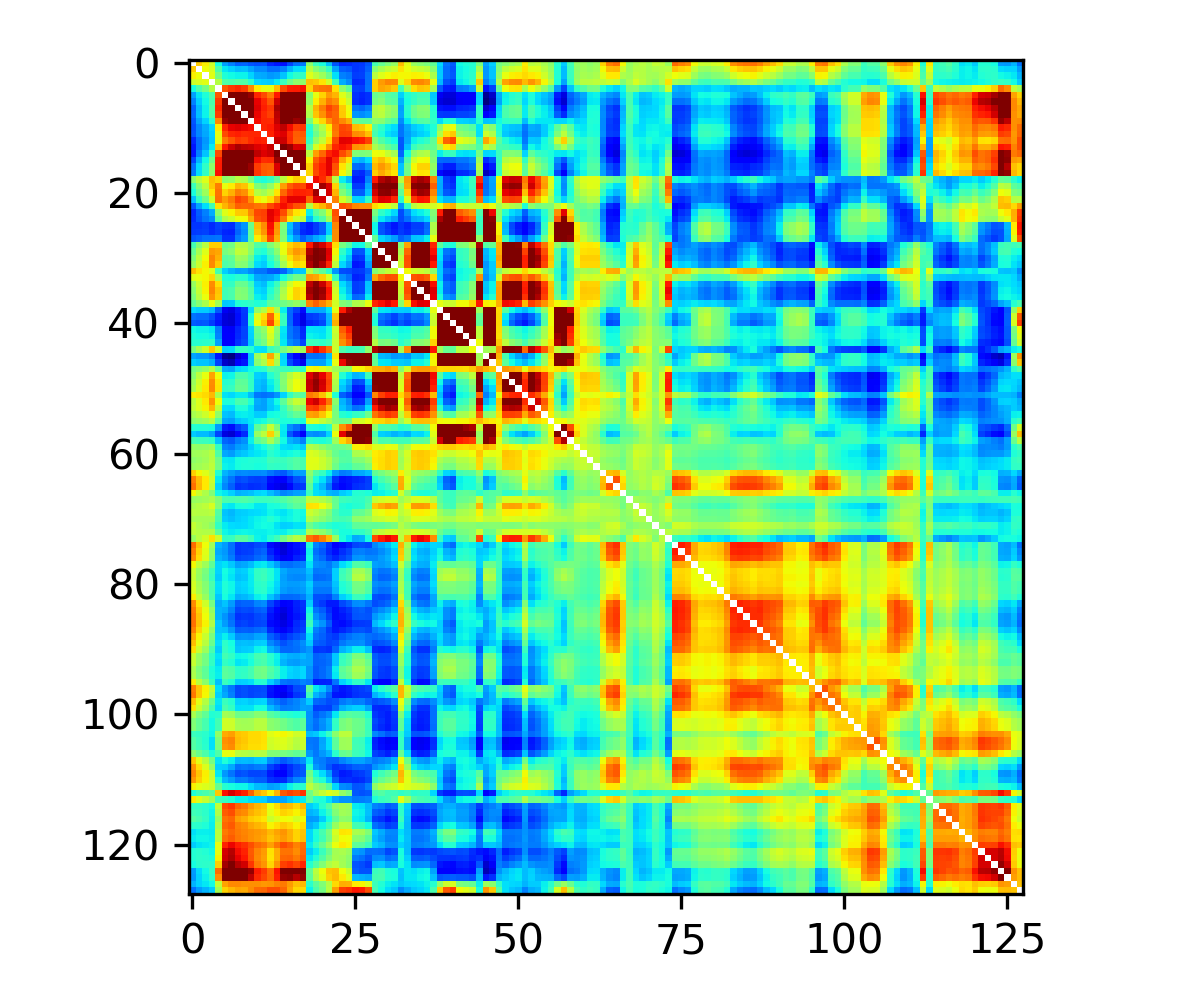}
 \end{subfigure}
 \begin{subfigure}[t]{0.32\textwidth}
 \centering
       \includegraphics[width=1\linewidth]{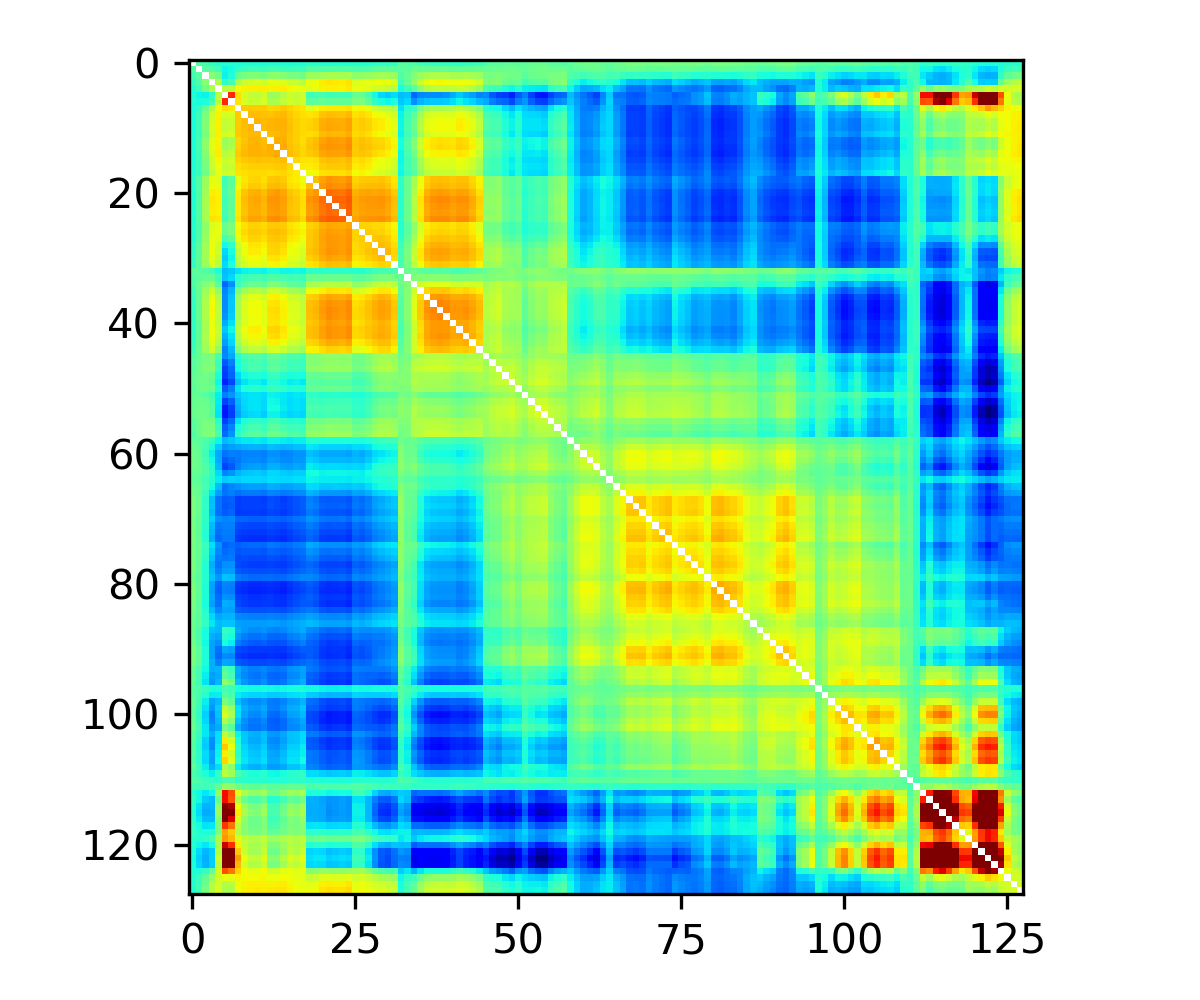}
 \end{subfigure}
\begin{subfigure}[t]{0.32\textwidth}
 \centering
       \includegraphics[width=1\linewidth]{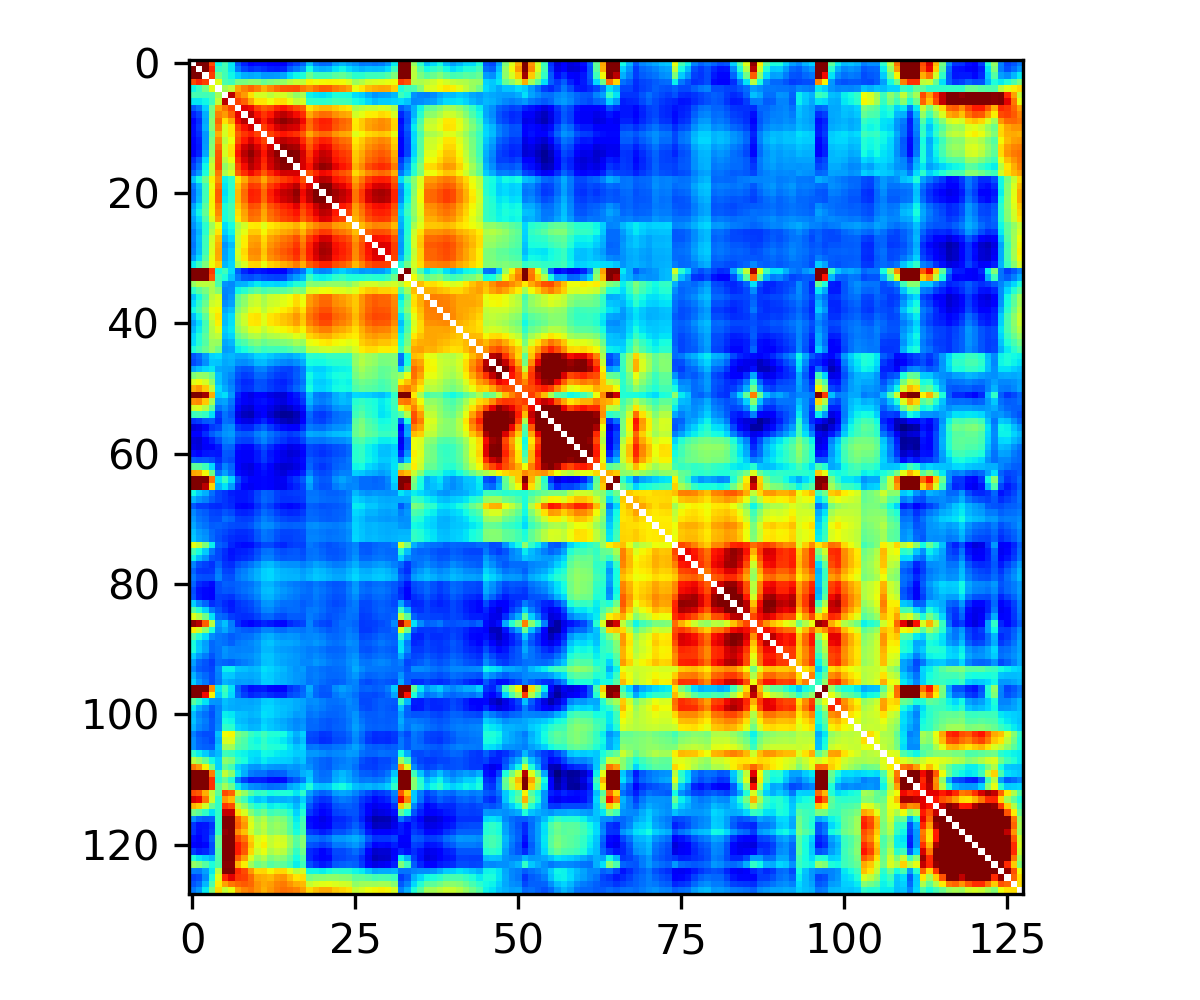}
 \end{subfigure}
  \caption{Fitted Laplacian shows the structure underneath the raw connectivity matrix.}
 \end{figure}

\end{document}